\documentclass[aps,twocolumn,showpacs,preprintnumbers,nofootinbib,prd,10pt,superscriptaddress]{revtex4-1}

\makeatletter
\def\l@subsubsection#1#2{}
\def\l@subsubsubsection#1#2{}
\makeatother

\usepackage{graphicx,amssymb,amsmath,amsthm,amsfonts,epsfig,epsf,fixmath}
\usepackage[usenames]{color}
\usepackage{epstopdf}

\usepackage{aas_macros}
\usepackage{bm}
\usepackage{dcolumn}
\usepackage{latexsym}
\usepackage{rotating}
\usepackage{longtable}

\usepackage{enumerate}
\usepackage{tensor,multirow}
\usepackage{url}
\usepackage[dvipsnames]{xcolor}
\usepackage[unicode]{hyperref}
\hypersetup{colorlinks=true, citecolor=MidnightBlue,
            linkcolor=MidnightBlue, urlcolor=MidnightBlue, linktocpage=true}

\newcommand{\tn}{\textnormal}

\newcommand{\ee}{\mathrm{e}}

\newcommand{\dd}{\mathrm{d}}
\newcommand*\DAlembert{\mathop{}\!\mathbin\Box}

\newcommand{\GSSI}{Gran Sasso Science Institute (GSSI), I-67100 L’Aquila, Italy}
\newcommand{\GranSasso}{INFN, Laboratori Nazionali del Gran Sasso, I-67100 Assergi, Italy}

\begin{document}
\title{Extreme mass-ratio inspirals as probes of scalar fields: \\ eccentric equatorial orbits around Kerr black holes}

\author{Susanna Barsanti}
\affiliation{Dipartimento di Fisica, ``Sapienza'' Universit\`a di Roma, Piazzale 
Aldo Moro 5, 00185, Roma, Italy}
\affiliation{Sezione INFN Roma1, Roma 00185, Italy}
\author{Nicola Franchini}
\affiliation{SISSA, Via Bonomea 265, 34136 Trieste, Italy and INFN Sezione di Trieste}
\affiliation{IFPU - Institute for Fundamental Physics of the Universe, Via Beirut 2, 34014 Trieste, Italy}
\author{Leonardo Gualtieri}
\affiliation{Dipartimento di Fisica, Universit\`a di Pisa, Largo B. Pontecorvo 3, 56127 Pisa, Italy}
\affiliation{INFN, Sezione di Pisa, Largo B. Pontecorvo 3, 56127 Pisa, Italy}
\author{Andrea Maselli}
\affiliation{\GSSI}
\affiliation{\GranSasso}
\author{Thomas P. Sotiriou}
\affiliation{School of Mathematical Sciences \& School of Physics and Astronomy, University of Nottingham, University Park, Nottingham, NG7 2RD, UK}
\affiliation{Nottingham Centre of Gravity, University of Nottingham, University Park, Nottingham, NG7 2RD, UK}

\begin{abstract} 
We study binary systems in which a stellar mass compact object spirals into a massive black hole, known as extreme mass ratio inspirals, in scenarios with a new fundamental scalar field. Earlier work has shown that, in most interesting such scenarios and to leading order in the mass ratio, the massive black holes can be adequately approximated by the Kerr metric and the imprint of the scalar field on the waveform is fully controlled by the scalar charge of the stellar mass object. Here we use this  drastic simplification in the inspiral modelling and consider eccentric equatorial orbits. 
We study how the scalar charge affects the orbital evolution for different eccentricities and different values of the black hole spin. We then determine how changes in the orbital evolution get imprinted on the waveform and assess LISA's capability to detect or constrain the scalar charge.

\end{abstract}

\maketitle

\section{Introduction}

In 2015, the landmark detection of gravitational waves (GWs)~\cite{LIGOScientific:2016aoc} paved the way to observations of strong-field gravity. The first searches for new fundamental physics in this regime by  the  interferometric detectors LIGO and Virgo~\cite{LIGOScientific:2019fpa} have found neither deviations from the General Relativity (GR), nor hints of new fundamental fields; still, these observations had relatively small signal-to-noise ratios (SNRs). The next generation of ground-based (like the Einstein Telescope~\cite{Sathyaprakash:2012jk}, recently included in the ESFRI Roadmap) and space-based (like the LISA mission~\cite{Audley:2017drz}, planned by ESA in 2037) detectors are expected to observe signals with SNRs of the order of one hundred, and hence they should be able to test GR in the strong-field regime of gravity with unprecedented accuracy.

New fundamental scalar fields are ubiquitous in cosmological models aimed to explain dark energy/matter components, in models of  quantum gravity, or in beyond-standard-model theories~\cite{Berti:2015itd,Barack:2018yly,Barausse:2020rsu}.
Extensions of GR can also be reformulated in such a way that modifications are encoded in additional fields that mediate the gravitational interaction~\cite{Sotiriou:2014yhm,Berti:2015itd,Barack:2018yly}. Scalar fields indeed provide the most straightforward way to include additional degrees of freedom within the gravity sector, in the so-called {\it scalar-tensor theories}.

Possible deviations from GR, or in the interactions between gravity and new fields that are part of some extension of the standard model, would be more likely to manifest in astrophysical environments characterized by relativistic velocities and high-curvature regimes, as those featured by the coalescence of compact binaries formed by neutron stars (NS) and black holes (BH). These systems, which are the main sources of gravitational waves for both ground and space interferometers, are therefore natural laboratories to test gravity in a genuine strong-field arena.

In this context extreme mass ratio inspirals (EMRIs), in which a stellar-mass compact object (SCO or secondary, with mass $m_{\rm p}$) inspirals into a massive black hole (MBH or primary, with mass $M\gg m_{\rm p}$), provide a special exception. Although merger rates are still uncertain~\cite{Babak:2017tow,Bonga:2019ycj}, EMRIs are expected to form due to the capture of SCOs by MBHs with masses in the range $\sim(10^4-10^9)M_\odot$, which are believed to reside in the centers of most galaxies, and to be surrounded by  nuclear star clusters~\cite{Ferrarese:2004qr}. Depending on the component masses, the final year before the plunge can lead to $\sim 10^4-10^5$ orbital cycles, most of which are accumulated when the secondary is in the region closer to the MBH. 
Such large number of orbits allows us to build a detailed map of the binary spacetime and to reconstruct with exquisite precision the source parameters~\cite{Babak:2017tow,Barack:2003fp,Babak:2006uv,Huerta:2011kt,Huerta:2011zi,Speri:2021psr,Piovano:2021iwv,Katz:2021yft,Maselli:2021men}.

EMRIs are expected to emit GWs at milli-Hertz frequencies, where LISA will be most sensitive. 
The accuracy of EMRI observations by LISA represents the key opportunity to explore a plethora of astrophysical phenomena~\cite{Seoane:2021kkk,Laghi:2021pqk,Berry:2019wgg,McGee:2018qwb,Amaro-Seoane:2007osp}, to study the environment in which EMRIs evolve~\cite{Annulli:2020ilw,Cardoso:2019rou,Hannuksela:2018izj,Barausse:2014tra,Yunes:2011ws,Kocsis:2011dr}, and to perform new and very sensitive tests of gravity or the nature of compact objects~\cite{Barausse:2020rsu,Yunes:2011aa,Barausse:2016eii,Blazquez-Salcedo:2016enn,Glampedakis:2005cf,Barack:2006pq,Cardoso:2018zhm,Cardoso:2018ptl,Datta:2019epe,Pani:2019cyc,Maggio:2021uge,Destounis:2020kss,Piovano:2020ooe,Destounis:2021mqv,Collodel:2021jwi,Sago:2021iku,Maselli:2020zgv,Maselli:2021men}.

One might be tempted to think that the detection of scalar fields by means of an EMRI may be hampered by no-hair theorems which state that, in several cases of scalar-tensor theories,  stationary BHs are described by the Kerr metric as in GR, with a vanishing (or constant) scalar field~\cite{1970CMaPh..19..276C, Bekenstein:1995un, Hawking:1972qk, Sotiriou:2011dz, Hui:2012qt}. Moreover, in the known exceptions of scalar-tensor theories for which no-hair theorems do not hold and stationary BHs can have non-trivial scalar field profiles, the scalar charge\,\footnote{Here the term ``scalar charge'' does not refer to a conserved, Noether charge. It only denotes the coefficient of $1/r$ in the far-field limit of the scalar field.} is controlled by the coupling between the scalar field and quadratic (or higher-order) curvature invariants~\cite{Campbell:1991kz,Mignemi:1992nt,Kanti:1995vq,Alexander:2009tp,Yunes:2011we,Kleihaus:2011tg,Herdeiro:2014goa,Sotiriou:2013qea,Sotiriou:2014pfa,Maselli:2015tta,Maselli:2015yva,Silva:2017uqg,Doneva:2017bvd,Antoniou:2017acq,Antoniou:2017hxj}. As a result, deviations from GR scale as inverse powers of the mass and are therefore strongly suppressed for massive BHs, which are then endowed with a negligible scalar charge.\footnote{Notable exceptions may include spin-induced BH scalarization \cite{Dima:2020yac,Herdeiro:2020wei,Berti:2020kgk} and superradiance \cite{Brito:2015oca}. In the later case, the scalar cloud is not infinitely long-lived however and the end state is a Kerr BH with lower spin. }  This naively appears to be a serious obstacle for searches of massless or very light scalar with LISA~\cite{Audley:2017drz}. It might in fact be a blessing in disguise, as pointed out in ~\cite{Maselli:2020zgv} (hereafter paper I).

The subtlety lies in the fact that, so long as the SCO of the EMRI, which has much smaller mass, carries some appreciable scalar charge, scalar emission will still be present and affect the orbital dynamics. Its effect will accumulate over the long time of observation and this can lead to significant deviation in the waveform. Hence, being able to approximate the spacetime around the MBH by a Kerr BH, on account of no-hair theorems or the scaling of the scalar charge with the mass, it turns out to provide an important simplification in waveform modelling. This simplification is applicable to a vast class of theories with a non-minimal coupling between the gravitational and an additional real, massless scalar field and has been employed in Paper I to develop a framework for modeling the EMRI dynamics and associated emission in a universal way, which only depends on the scalar charge of the SCO. The effect of the scalar charge is to induce an extra loss of energy and angular momentum which accelerates the coalescence leaving an imprint on the GW signal detectable by LISA. An assessment of the capability of LISA to measure the scalar charge of the secondary, with an estimate of the expected errors, has been performed in~\cite{Maselli:2021men} (hereafter paper II).

Paper I takes into account only equatorial and circular orbits around a non-rotating MBH, while Paper II included the MBH rotation for a single value of the spin, $a=0.9\,M$.
Here we provide a substantial extension of the framework to a more astrophysically motivated scenario, by studying in detail how the spin of the primary and the eccentricity affect the scalar (and the gravitational) emission of the binary and its phase evolution. The inclusion of eccentricity is particularly relevant for the long orbits for EMRIs. The orbit can have a large initial eccentricity, in which case the binary emits bursts of GWs at each pericenter passage, before evolving into a more circular inspiral. 
We remark that we do not make any assumption on the nature of the secondary: it can be either a BH or a NS. We neglect the spin of the secondary, since we are only interested in the leading-order corrections to the energy and momentum fluxes and then to the GW signal~\cite{Piovano:2020zin,Piovano:2021iwv}.

Whenever the MBH is adequately described by the Kerr metric, as per the assumptions of Papers I and II and our analysis below, and since the mass ratio of the binary $q=m_{\rm p}/M$ is very small, 
the inspiral of the SCO and the resulting GW emission can be studied using the well-known perturbation theory around Kerr spacetime, pioneered by Teukolsky~\cite{Teukolsky:1973ha} (see also~\cite{chandrasekhar1998mathematical}).
In this paper we consider leading-order, ``adiabatic'', perturbations, which lead to $\mathcal{O}(q^{-1})$ 
terms within the gravitational wave phase, 
neglecting higher-order corrections on the mass 
ratio.

In Section~\ref{sec:assumptions} we review the theoretical framework, showing that in a wide class of gravity theories the primary of an EMRI can be described by a Kerr BH, and deriving the field equations for the metric and the scalar field. We also discuss the description of eccentric, equatorial orbits in Kerr spacetime. In Section~\ref{sec:scalarkerr} we describe our approach to  solve the field equations to compute the energy and angular momentum fluxes,  the dephasing of the gravitational waveform due to the scalar field emission, and the faithfulness between GW signals with and without the scalar charge. In Section~\ref{sec:results} we present the results of our numerical integration, discussing how the fluxes and the dephasing depend on the orbital parameters of the EMRI, and assessing the detectability of the scalar field by LISA. Finally, in Sec.~\ref{sec:conclusions} we draw our conclusions.

\section{Theoretical set up}\label{sec:assumptions}
In this section we  review the theoretical background behind our approach, and provide the relevant formalism necessary to compute the GW flux emitted by EMRIs with extra scalar charges. We use geometric units with $c=G=1$.

\subsection{Action}\label{sec:action}

We consider theories in which a scalar field $\varphi$ is potentially non-minimally coupled to the spacetime metric ${\bf g}$. We assume $\varphi$ to 
be massless. We expect that our result will be valid, qualitatively, for very light scalars as well, while more massive scalar fields would be significantly suppressed at large distances, decreasing deviations from GR at the level of the waveforms. We will discuss this in detail in a forthcoming paper \cite{papermassive}.
The most general action that can describe such theories is (see paper I): 
\begin{equation}
    S\left[\textbf{g}, \varphi, \Psi \right] = S_0\left[\textbf{g}, \varphi\right] + \alpha S_{\rm c} \left[\textbf{g}, \varphi\right] + S_{\rm m}\left[\textbf{g}, \varphi, \Psi\right]\ .
    \label{action}
\end{equation}
$S_0$ includes the Einstein-Hilbert action and the canonical 
kinetic term for the scalar field: 
\begin{equation}
    S_0 = \int \dd^4 x \frac{\sqrt{-g}}{16 \pi} \left(R - \frac{1}{2} \partial_\mu \varphi \partial^{\mu} \varphi \right)\ ,
\end{equation}
where $R$ is the Ricci scalar. The non-minimal coupling between ${\bf g}$ and $\varphi$ is encoded in $\alpha S_{\rm c}$, with $\alpha$ being a constant of dimensions $\left[\alpha\right] = (\text{mass})^{n}$. We assume $S_{\rm c}$ to be analytic in $\varphi$. Finally, matter fields are described by $S_{\rm m}$. 

For EMRIs the typical scale of the secondary is much smaller than the 
characteristic scale of the ``exterior spacetime'', {\it i.e.}~of the solution of the field equations in the absence of the small body. Therefore, we use the {\it skeletonized approach} developed in~\cite{1975ApJ...196L..59E,Damour:1992we,Julie:2017ucp, Julie:2017rpw}, in which the inspiraling body is treated as a point particle and $S_{\rm m}$ is replaced by the particle action $S_{\rm p}$ given by the integral of a scalar function $m(\varphi)$ over the worldline followed by the secondary, $y^{\mu}_{\rm p}(\lambda)$ (in a reference frame $\{y^\mu\}$): 
\begin{equation}
    S_{\rm p} = - \int m\left(\varphi\right) \dd s = - 
    \int m \left(\varphi \right) \sqrt{g_{\mu \nu} \frac{\dd y^\mu_{\rm p}}{\dd\lambda}\frac{\dd y^\nu_{\rm p}}{\dd\lambda}} 
    \dd\lambda\ .\label{eq:partac}
\end{equation}
The function $m(\varphi)$ depends on the value of the scalar field at the location of the particle, and accounts for the coupling of the body to its scalar field environment. This approximation holds at linear order in the mass ratio.

The modelling of the exterior spacetime is greatly simplified in  theories that belong in the following two classes:
\begin{enumerate}
\item The theory satisfies a no-hair theorem~\cite{1970CMaPh..19..276C, Bekenstein:1995un, Hawking:1972qk, Sotiriou:2011dz, Hui:2012qt}. This case covers several classes of scalar-tensor theories (including those in which $\alpha=0$ and the scalar field couples to matter). 
\item The theory evades the no-hair theorems but has a dimensionful coupling constant $\alpha$, with $n \geq 1$; in this case we also assume that the BH solutions  are  continuously connected to the corresponding solution in GR as $\alpha \rightarrow 0$. All known examples of theories that allow for scalar hair fall in this class so far, {\em e.g.}~scalar-Gauss--Bonnet gravity and Chern--Simons gravity~\cite{Mignemi:1992nt,Kanti:1995vq,Sotiriou:2013qea,Kleihaus:2011tg,Alexander:2009tp}.
\end{enumerate}
In case 1 it is clear that stationary BHs are described by the Kerr metric. In case 2, the Kerr metric serves as an excellent approximation at low orders in the mass ratio.
In case 2 the BH spacetime is continuously connected to the Kerr solution as $\alpha \rightarrow 0$, and since the only dimensionful scale of the Kerr metric is the mass $M$ of the MBH, any correction to the latter must depend on the dimensionless parameter $\zeta$ defined as: 
   \begin{equation}
    \zeta = \frac{\alpha}{M^n} = q^n \frac{\alpha}{m_{\rm p}^n}=q^n \zeta_{\rm p}, 
    \end{equation} 
where $q \equiv m_{\rm p}/M$ is the binary mass ratio, which for EMRIs is $q\ll 1$, and $\zeta_{\rm p}=\alpha / m_{\rm p} ^n$. Bounds on $\alpha$ obtained from astrophysical observations imply $\zeta_{\rm p}< 1$~\cite{Nair:2019iur}. Therefore, the parameter which controls deviations from the Kerr geometry is also small,  $\zeta \ll 1$, actually suppressed by powers of the mass ratio. Note that there can also be theories that are in neither Case 1 or Case 2 above, but for specific EMRIs the massive BH can be well approximated by Kerr ({\em e.g.}~it just happens to carry a negligible scalar charge).

We can conclude that in an EMRI, for a wide class of  theories with a scalar field the spacetime of the primary is given by the Kerr metric to order $\mathcal{O}(q^n\zeta_{\rm p})$ in the mass ratio. The secondary moves in this background and its motion and radiation emission can be studied using standard perturbation theory on the Kerr spacetime.

We remark that a theory in which the primary object of an EMRI 
is not described with good accuracy by the Kerr metric would exhibit larger deviations from GR in the gravitational waveform  than those found in this article. Hence, our results can be considered as a conservative estimate.

\subsection{Field equations}\label{fields}

The field equations are obtained by varying action \eqref{action} (with $S_{\rm m}$ now replaced by $S_{\rm p}$) with respect to the fields. Variation with respect to  the metric gives the modified Einstein equations~\cite{Maselli:2020zgv}: 
\begin{equation}
    G_{\mu\nu} = R_{\mu\nu} - \frac{1}{2} g_{\mu\nu}R = 8 \pi T^{\rm scal}_{\mu \nu}+ \alpha T^c_{\mu\nu}+ T^p_{\mu\nu}\,,\label{eq:modein}
\end{equation}
where $R_{\mu \nu}$ is  the Ricci tensor, $R$ is the Ricci scalar,
\begin{equation}
    T^{\rm scal}_{\mu \nu} = \frac{1}{16\pi}\left[\partial_{\mu} \varphi\partial_{\nu} \varphi - \frac{1}{2}g_{\mu\nu}\left(\partial \varphi\right)^2\right]
    \label{tensorscalar}
\end{equation}
is the stress-energy tensor of the scalar field and 
\begin{equation}
    T^{\rm c}_{\mu \nu} = - \frac{16 \pi}{\sqrt{-g}}\frac{\delta S_{\rm c}}{\delta g^{\mu \nu}} 
\end{equation}
is the stress-energy tensor associated to the coupling between the scalar and the gravitational fields. 
The stress energy tensor of the secondary body obtained by varying the skeletonized action  $S_{\rm p}$~\eqref{eq:partac} is given by

\begin{equation}\label{eq:stressPP}
    T^{{\rm p}\, \alpha \beta}  =
    8 \pi \int m(\varphi) \frac{\delta^{(4)}(x- y_{\rm p}(\lambda))}{\sqrt{-g}} \frac{\dd y_{\rm p}^{\alpha}}{\dd\lambda}\frac{\dd y_{\rm p}^\beta}{\dd\lambda}  \dd\lambda\  . 
\end{equation}

We describe this system using perturbation theory with respect to the mass ratio $q\ll1$. We expand the scalar field as $\varphi = \varphi_0+\varphi_1$, where  $\varphi_0$ is the constant background field, and $\varphi_1$ is  the perturbation induced by the secondary.  

We can now show that within our approach, both $T^{\rm scal}_{\mu \nu}$  and $\alpha T^{\rm c}_{\mu \nu}$ can be neglected at  leading (adiabatic) order in $q$, as they contribute only to higher (post-adiabatic) orders in the small-ratio expansion.
Indeed, since the background scalar field $\varphi_0$ is constant, the stress-energy tensor of the scalar field $T^{\rm scal}_{\mu \nu}$ is quadratic in the perturbation $ \varphi_1$, {\it i.e.}~$\mathcal{O}(q^2)$ [or $\mathcal{O}(q)$ with respect to the leading contribution to Eq.~\eqref{eq:modein}, which is given by the stress-energy tensor of the secondary].
Moreover, since $\left[S_0\right]= (\text{mass})^{2}$, $\left[S_c\right]= (\text{mass})^{2-n}$ and in an EMRI $S_c$ is evaluated on the background of the MBH, where the only dimensionful scale is its mass $M$, we expect that: 
\begin{equation}
    S_c \sim M^{-n} S_0.
    \label{approx}
\end{equation}
Thus, 
\begin{equation}
    \alpha T^c_{\mu \nu} = - \frac{16 \pi \alpha }{\sqrt{-g}}\frac{\delta S_c}{\delta g^{\mu \nu}} \sim  - \frac{16 \pi \alpha M^{-n} }{\sqrt{-g}}\frac{\delta S_0}{\delta g^{\mu \nu}},
\end{equation}
and, since $\alpha M^{-n} = \zeta \ll 1$,
\begin{equation}
    \alpha T^c_{\mu \nu} \sim \zeta G_{\mu \nu} \ll G_{\mu \nu}\ .
\end{equation}
Therefore $\alpha T^c_{\mu \nu}$ is $\mathcal{O}(q^n\zeta_{\rm p})$ with respect to the Einstein tensor and can be neglected. The field equations for the gravitational field are then:
\begin{equation}
    G^{\alpha \beta} = 8 \pi  \int m(\varphi) \frac{\delta^{(4)}\left(x-y_p(\lambda)\right)}{\sqrt{-g}}\frac{\dd y^{\alpha}_p}{\dd\lambda} \frac{\dd y^{\beta}_p}{\dd\lambda} \dd\lambda\ .
\label{graveq}
\end{equation}

Variation of the action~\eqref{action} with respect to the scalar field gives:
\begin{equation}
    \DAlembert \varphi + \frac{16 \pi \alpha}{\sqrt{-g}}\frac{\delta S_c}{\delta \varphi} = 16 \pi \int m'\left(\varphi\right) \frac{\delta ^{(4)}\left(x - y_p (\lambda)\right)}{\sqrt{-g}}\dd\lambda\ ,\label{eq:eqsc0}
\end{equation}
where $m'(\varphi)= dm(\varphi)/d\varphi$. 

The scaling property~\eqref{approx} implies that the second term on the left-hand side of Eq.~\eqref{eq:eqsc0} can be neglected:
\begin{equation}
    \frac{16 \pi \alpha}{\sqrt{-g}}\frac{\delta S_c}{\delta \varphi} \sim  \frac{16 \pi \alpha M^{-n}}{\sqrt{-g}}\frac{\delta S_0}{\delta \varphi} \sim \zeta \DAlembert \varphi \ll \DAlembert \varphi\ . 
\end{equation}
Thus, the scalar field equation reduces to: 
\begin{equation}
    \DAlembert \varphi = 16 \pi \int m'\left(\varphi\right) \frac{\delta ^{(4)}\left(x - y_p (\lambda)\right)}{\sqrt{-g}}\dd\lambda\ . 
    \label{scaleq}
\end{equation}
The functions $m(\varphi)$ and $m'(\varphi)$ are evaluated at the value of the scalar field in the location of the particle, namely $\varphi_0$, and are determined uniquely by the properties of the secondary.

Let us consider a reference frame $\{\tilde{x}_{\mu}\}$ centered on the latter. The solution to the scalar field equation in a buffer region inside the world-tube of the inspiralling body, still far enough to have a metric which can be written as a flat spacetime perturbation, can be written as: 
\begin{equation}
    \varphi = \varphi_0 + \frac{m_{\rm p} d}{\tilde{r}} + O\left(\frac{m_{\rm p}^2}{\tilde{r}^2}\right), \label{asymtpsolphi}
\end{equation}
where $d$ is the dimensionless scalar charge of the body. 
Matching the solution in the buffer region by replacing Eq.~\eqref{asymtpsolphi} into Eq.~\eqref{scaleq}, we obtain a relation between the charge and the mass function:
\begin{equation}
    \frac{m'(\varphi_0)}{m_{\rm p}} = - \frac{d}{4}\ .
    \label{mprimo}
\end{equation}
Since in the weak-field limit the $(tt)$-component of the particle's stress energy tensor, given by 
\begin{equation}
    T^{{\rm p}\,tt} = 8 \pi m(\varphi_0) \delta^{(3)}\left(x^i - y^i_p(\lambda)\right)+ O \left( \frac{m_p}{\tilde{r}}\right)\ ,
\end{equation}
reduces to the matter density of the particle,
\begin{equation}
    \rho = m_{\rm p} \delta^{(3)}\left(x^i - y^i_{\rm p}(\lambda)\right)\ , 
\end{equation}
we obtain
\begin{equation}
    m(\varphi_0) = m_{\rm p}\ . 
    \label{m}
\end{equation}
Using Eqns.~\eqref{mprimo} and \eqref{m} for $m(\varphi)$ we finally obtain the following expressions for the field equations  (to first order in the mass ratio):
\begin{equation}
    G^{\alpha \beta} = 8 \pi m_{\rm p}  \int \frac{\delta^{(4)}\left(x-y_p(\lambda)\right)}{\sqrt{-g}}\frac{\dd y^{\alpha}_p}{\dd\lambda} \frac{\dd y^{\beta}_p}{\dd\lambda} \dd\lambda \ ,
\label{einstein}
\end{equation}
and
\begin{equation}
    \DAlembert \varphi = - 4 \pi d m_{\rm p} \int \frac{\delta^{(4)} \left(x - y_p(\lambda)\right) }{\sqrt{-g}}\dd\lambda\ . 
    \label{scal_schw}
\end{equation}
Eqs.~\eqref{einstein}, \eqref{scal_schw} are the key ingredients of our approach, and lead to fundamental results.  The gravitational field equations, Eqs.~\eqref{einstein}, coincide with those of the GR case.  The scalar field equation~\eqref{scal_schw}, instead, has a source term whose magnitude is controlled by the dimensionless scalar charge carried by the secondary. Therefore, all changes in the EMRI evolution given by the extra scalar field are uniquely and universally specified by the value of $d$. 
For many gravity theories, the latter can be uniquely mapped 
to the theoretical parameters which control deviations from 
GR. In such cases, future measurements of $d$ with LISA 
observations, can be translated to constraints on the 
fundamental parameters that characterize beyond GR theories~\cite{Maselli:2021men,Julie:2022huo}.

These results have important consequences.
From a computational standpoint, the problem is reduced to the resolution of the same equations as in GR, plus a single extra equation describing the scalar field on a Kerr background. Moreover, the universality of Eq.~\eqref{scal_schw} allows for tests of GR without any assumptions regarding the origin of the deviations (see Papers I and II).

\section{Adiabatic inspiral}
\label{sec:scalarkerr}
As discussed in Sec.~\ref{sec:assumptions}, we solve the 
modified Einstein field equations \eqref{einstein}-\eqref{scal_schw} 
using a perturbative approach, at leading order in the mass ratio $q$.
Therefore, we describe the EMRI inspiral within the adiabatic approximation, in which the timescale of the energy and angular momentum dissipation is much larger than the dynamical timescale, and thus the secondary object follows a sequence of geodesics until the plunge. We describe the equatorial 
eccentric motion of the body in Appendix~\ref{sec:geodesics}. In the following, we discuss how to compute the energy and angular momentum fluxes to determine the orbital evolution for a prograde equatorial eccentric inspiral.

\subsection{Scalar field perturbations}
By decomposing the metric and scalar field functions using the Newman-Penrose formalism we obtain a single master equation for both tensor and scalar perturbations, the Teukolsky equation~\cite{Teukolsky:1973ha}:

\begin{widetext}
\begin{multline}
    \left[\frac{\left(r^2 + a^2\right)^2}{\Delta} - a^2 \sin{\theta}^2 \right] \partial^2_0 \psi^{(s)} + \frac{4aMr}{\Delta} \partial_0 \partial_{\phi}\psi^{(s)} + \left[\frac{a^2}{\Delta} - \frac{1}{\sin{\theta}^2} \right] \partial^2_{\phi} \psi^{(s)} 
    - \Delta^{-s} \partial_r \left( \Delta^{s+1} \partial_r \psi^{(s)} \right) - \frac{1}{\sin{\theta}} \partial_{\theta} \left( \sin{\theta} \partial_{\theta} \psi^{(s)} \right)
     \\ - 2s \left[\frac{a \left(r-M \right)}{\Delta}+i \frac{\cos{\theta}}{\sin{\theta}^2} \right] \partial_{\phi} \psi^{(s)}  -2s \left[ \frac{2M(r^2-a^2)}{2\Delta} - r - ia \cos{\theta} \right] \partial_0 \psi^{(s)} 
    + (s^2 \cot{\theta}^2 -s ) \psi^{(s)} = 4 \pi \Sigma T^{(s)}\ .
    \label{teukeq}
\end{multline}
\end{widetext}

The index $s$ is the {\it spin weight} of the perturbation, such that $s=0$ for a scalar field, $s=\pm1$ for vector perturbations, $s=\pm2$ for gravitational perturbations (see Appendix~\ref{sec:geodesics} for notations and conventions on the Kerr metric):
\begin{equation}
    \psi^{(-2)} = \left( r - i a \cos{\theta} \right)^4 \Psi_4\ , \qquad \psi^{(0)} = \varphi\ , 
    \label{psipsi4}
\end{equation}
where $\Psi_4$ is one of the Weyl scalars, and the source term $T$ is a combination of the components of the stress-energy tensor.

The Teukolsky equation~\eqref{teukeq} decouples into an angular and a radial component if we apply a Fourier transform on $\psi^{(s)}$ and $T^{(s)}$ and expand them in spin-weighted spheroidal harmonics $S^{(s)}_{\ell m}(\theta,\omega)$:
\begin{equation}
    \psi^{(s)} (t, r,\theta, \phi) = \int \dd\omega \sum_{\ell m}{ \tilde{R}^{(s)}_{\ell m}(r,\omega) S^{(s)}_{\ell m}(\theta,\omega) e^{i m \phi} e^{-i\omega t}}\ ,
    \label{psisdec}
\end{equation}
\begin{equation}
    4 \pi \Sigma T^{(s)} = \int \dd\omega \sum_{\ell m}{ \tilde{J}^{(s)}_{\ell m}(r,\omega) S^{(s)}_{\ell m}(\theta,\omega) e^{i m \phi} e^{-i\omega t}}\ .
    \label{sourcedec}
\end{equation} 
The  spheroidal harmonics $S^{(s)}_{\ell m}$ are solutions of the equation:
 \begin{multline}
    \left[\frac{1}{\sin\theta} \frac{\dd}{\dd\theta}  \left( \sin\theta \frac{\dd}{\dd\theta}\right) - \gamma^2 \sin^2\theta - \frac{(m+s\cos\theta)^2}{\sin^2\theta} \right.\\- 2\gamma s \cos\theta + s+ 2m\gamma + {}_s\lambda_{\ell m}\bigg] S^{(s)}_{\ell m} (\theta,\gamma) = 0\ ,\label{eq:spherharm}
\end{multline}
where $\gamma = a \omega$, and the parameter ${}_s\lambda_{\ell m}$ is the {\it angular eigenvalue}, determined by solving Eq.~\eqref{eq:spherharm}. The spheroidal harmonics satisfy the orthogonality relation
\begin{equation}
    \int  S^{(s)}_{\ell m}(\theta,\gamma) e^{im\phi} 
    S^{(s)\,*}_{\ell' m'}(\theta,\gamma) e^{-im'\phi} 
    \dd\Omega =\delta_{\ell \ell'}\delta_{mm'}\,. 
    \label{ortogonality}
\end{equation}
The radial components $\tilde{R}^{(s)}_{\ell m}(r,\omega)$ satisfy the following equation
\begin{align}
\Delta^{-s} \frac{\dd}{\dd r} \bigg[&\Delta^{s+1}\frac{\dd \tilde{R}^{(s)}_{\ell m}}{\dd r}\bigg] + \bigg[\frac{K^2 - 2 i s (r-M)K}{\Delta}\nonumber\\
&+ 4 i s \omega r - {}_s\lambda_{\ell m} \bigg]\tilde{R}^{(s)}_{\ell m} = 
\tilde{J}^{(s)}_{\ell m}\ , 
\label{eq:radial}
\end{align}
where $K=(r^2 + a^2)\omega - a m$. Since gravitational perturbations have been extensively studied in literature we refer the reader to Appendix~\ref{appendix:gravitational_perturbations} for further details. Hereafter,  we focus on the scalar sector, $s=0$ and suppress the index associated with the spin weight $s$ to lighten the notation.

Eq.~\eqref{eq:radial} can be solved using standard 
Green-function techniques.
For $s=0$, it becomes
\begin{equation}
   \frac{\dd}{\dd r}\left(\Delta \frac{\dd\tilde{R}_{\ell m}(r,\omega)}{\dd r}\right) + V(r) \tilde{R}_{\ell m}(r,\omega)  = \tilde{J}_{\ell m} \ , 
  \label{masterR}
\end{equation}
where $V(r) =(K^2/\Delta)- \lambda$.
Hereafter, unless specified differently, we will drop the 
multipolar indices $(\ell,m)$. To solve Eq.~\eqref{masterR} we define the auxiliary function 
\begin{equation}
    Y(\omega,r)\equiv \sqrt{r^2+a^2} \tilde{R}(\omega,r)\,.
    \label{y}
\end{equation}
Substituting $Y(\omega,r)$ 
in Eq.~\eqref{masterR} the radial equation becomes
\begin{equation}
    \frac{\dd^2}{\dd r_\star^2} Y +  V(\omega) Y = J_Y\ ,
    \label{masterscalk}
\end{equation}
where the potential $V(\omega)$ and the source term 
$J_Y$ are given by
\begin{align}
    V(\omega) =& \frac{K^2 - \lambda \Delta}{(r^2+a^2)^2}-G^2 - \frac{\dd G}{\dd r_{\star}}\ ,\nonumber\\ 
    J_Y =& \tilde{J} \frac{\Delta}{(a^2+r^2)^{3/2}}\ ,
    \label{eq:jy}
\end{align}
with $G = r \Delta / (r^2+a^2)^2$ and $r_\star$ is the tortoise coordinate defined by 
\begin{equation}
 \frac{\dd r_\star}{\dd r} = \frac{r^2 + a^2}{\Delta}\ .
 \label{tortoisekerr}
 \end{equation}

The homogeneous solution for Eq.~\eqref{masterscalk} 
is found by specifying proper boundary conditions at 
the horizon and at infinity where, according to the 
properties of the potential $V(\omega)$~\cite{Teukolsky:1973ha}
\begin{equation}
\begin{cases}
     \frac{\dd^2}{\dd r_\star^2} Y + k^2 Y = J_Y &\text{for $r \rightarrow r_{+}$}\ ,\\
     \frac{\dd^2}{\dd r_\star^2} Y +\omega^2  Y = J_Y &\text{for $r \rightarrow \infty$}\ ,
\end{cases}
\end{equation}
where $k = \omega - m \Omega_+$ and $\Omega_+ = \frac{a}{2Mr_+}$. 
The homogeneous solutions $Y_-/Y_+$ which satisfy  
the condition of purely ingoing/outgoing wave at the 
horizon/infinity have the following asymptotic behaviour: 
\begin{equation}
\begin{cases}
    Y_- = e^{-ikr_\star} &\text{for $r \rightarrow r_{+}$}\ ,\\
    Y_- = A_\textnormal{in} e^{-i\omega r_\star} + A_\textnormal{out} e^{i\omega r_\star} &\text{for $r \rightarrow \infty$}\ ,
\end{cases}
\end{equation}

\begin{equation}
\begin{cases}
    Y_+ = B_\textnormal{in} e^{-ikr_\star}+ B_\textnormal{out} e^{i k r_\star} &\text{for $r \rightarrow r_{+}$}\ ,\\
    Y_+ = e^{i\omega r_\star} &\text{for $r \rightarrow \infty$}\,.
\end{cases}
\end{equation}
The general solution for $Y(\omega,r)$ is then obtained 
by integrating the former over $J_Y$:
\begin{equation}
    Y = Y_+ \int^{r_{\star}}_{-\infty} \frac{Y_- J_Y \dd r_\star}{W_Y}+ Y_- \int^{+\infty}_{r_{\star}} \frac{Y_+ J_Y \dd r_\star}{W_Y}\ ,\label{fullY}
\end{equation}
where $W_Y= Y'_+ Y_- - Y_+ Y'_-,$ is the 
Wronskian and primes denote derivatives with respect 
to $r_{\star}$. 
From Eq.~\eqref{y} we also obtain the asymptotic behavior of 
$\tilde{R}_{-,+}$:
\begin{equation}
\begin{cases}
    \tilde{R}_- = e^{-ikr_\star} &\text{for $r \rightarrow r_{+}$}\ ,\\
    \tilde{R}_- = \frac{A_\textnormal{in}}{r} e^{-i\omega r_\star} + \frac{A_\textnormal{out}}{r} e^{i\omega r_\star} &\text{for $r \rightarrow \infty$}\ ,
\end{cases}
\end{equation}
\begin{equation}
\begin{cases}
    \tilde{R}_+ = B_\textnormal{in} e^{-ikr_\star}+ B_\textnormal{out} e^{i k r_\star} &\text{for $r \rightarrow r_{+}$}\ ,\\
    \tilde{R}_+ = \frac{e^{i\omega r_\star} }{r} &\text{for $r \rightarrow \infty$}\ .
\end{cases}
\end{equation}
A general solution for $\tilde{R}$ can be constructed as in 
\eqref{fullY}.

\subsection{Source terms}
We shall now derive the final form of the general solution. Let's first define the integral of the 
homogeneous solution over the source term:
\begin{equation}
   \delta \varphi^{-,+}_{\ell m\omega} = \int^{+\infty}_{-\infty} \frac{Y_{+,-} J_{Y} \dd r_\star}{W_{Y}}.
   \label{deltaphi}
\end{equation} 
Projecting the source term $J$ on the spheroidal functions we find 
the explicit expression of $\tilde{J}$ 
\begin{align}
    \tilde{J}_{\ell m}
    &=-2d\int^{+\infty}_{-\infty} \frac{m_p \delta[r-r_p(t)]}{\dot{t}} S^*e^{i[\omega t-m\phi_p(t)]}\dd t\ , 
\end{align}
where $S^*$ has to be evaluated on the equatorial 
plane at $\theta=\pi/2$. Replacing the former into Eq.~\eqref{eq:jy} and 
\eqref{deltaphi} and integrating over the radial coordinate we obtain 
\begin{equation}
    \delta \varphi_{\ell m\omega}^{-,+} = \int^{+\infty}_{-\infty} I^{-,+}\left[r_p(t)\right]e^{i[\omega t -m\phi_p(t)]}dt, 
    \label{dphiI}
\end{equation}
with 
\begin{equation}
   I^{-,+}\left[r_p(t)\right] = \left[\frac{Y_{-,+}}{W_Y} \frac{-2dm_p}{(a^2+r^2)^{1/2}} \frac{ S^*}{\dot{t}}\right]_{r=r_p(t)} \,.
    \label{defI}
\end{equation}
Finally, we define the functions 
\begin{equation}
    \alpha(t)^{-,+} = I^{-,+}[r(t)] e^{-im[\phi(t)-\Omega_\phi t]}\,. 
    \label{eq:alpha}
\end{equation}
They are periodic in $r$ with period $T_r$, and thus 
they can be expanded as a Fourier series as
\begin{equation}
    \alpha(t)^{-,+} = \sum^{+\infty}_{n=-\infty} \hat{\alpha}^{-,+}_n e^{-in\Omega_r t}\ . 
\end{equation}
The frequencies $\Omega_i$ with $i = (\phi,r)$ are defined in Appendix \ref{sec:geodesics}, eq. \ref{eq:frequencies_def}. Replacing the expression of $I[r_p(t)]$ in terms of the Fourier 
expansion of $\alpha(t)$ in Eq.~\eqref{dphiI} and 
performing the time integral we finally obtain
\begin{equation}
    \delta \varphi^{-,+}_{\ell m\omega} =  \sum^{+\infty}_{n=-\infty} 
    \delta \hat{\varphi}^{-,+}_{\ell mn} \delta(\omega -\omega_{mn})\ ,
    \label{eq:deltaomega}
\end{equation}
where $\omega_{mn}=m\Omega_\phi+n\Omega_r$. The coefficients 
$\delta \hat{\varphi}^{-,+}_{\ell mn}=2\pi \alpha_n$ are given 
by the integral
\begin{equation}
    \delta \hat{\varphi}^{-,+}_{\ell mn} = \frac{2 \pi}{T_r} \int^{T_r}_0 
    \alpha(t) e^{i n \Omega_r t} \dd t\ .
\end{equation}
Substituting the expression~\eqref{eq:alpha} for $\alpha(t)$ and 
changing the integration variable from $t$ to $\chi$ (see Appendix~\ref{sec:geodesics}) yields:
\begin{align}
   \delta \hat{\varphi}^{-,+}_{\ell mn} =& \Omega_r  \int^{2 \pi}_0  \dd\chi \frac{V_t(\chi)I^{-,+}(\chi)}{J(\chi)\sqrt{V_r(\chi)}}e^{i[ \omega_{mn}t(\chi)-m\phi(\chi)]}\nonumber\\
   =& \Omega_r  \int^{ \pi}_0  \dd\chi \frac{V_t(\chi)I^{-,+}(\chi)}{J(\chi)\sqrt{V_r(\chi)}} 
    [e^{i[\omega_{mn}t(\chi)-m\phi(\chi)]}\nonumber \\ 
   &\phantom{aaaaaaaaaaaa} + e^{-i\omega_{mn}t(\chi)+im\phi(\chi)}]\ .  \label{amp}
\end{align}

The amplitudes $\delta \varphi^{-,+}_{\ell mn}$ computed through Eq.~\eqref{amp} are needed to compute the energy and angular momentum fluxes emitted by the binary. 

\subsection{Energy and angular momentum fluxes of the scalar field}
The energy flux associated to the scalar field can be derived through the effective stress-energy tensor for $\varphi$, as previously done {\em e.g.} in \cite{Warburton:2010eq}, which coincides with the Isaacson tensor in GR~\cite{Stein:2010pn,Blazquez-Salcedo:2016enn}. 
The stress-energy tensor for the scalar field is given in Eq.~\eqref{tensorscalar} and the energy flux reads: 
\begin{equation}
    \dot{E}^{\pm}_{\rm scal}=\frac{\dd E^{\pm}_{\rm scal}}{\dd t} =  \mp  \Delta \int T^{\rm scal}_{tr}\dd\Omega \ , 
\end{equation}
where 
$T^{\rm scal}_{tr} = (16 \pi)^{-1} \varphi_{,t} \varphi^*_{,r}$ and the upper (lower) sign indicates the emission at infinity (horizon).
The time derivative of $\varphi = \sum_{\ell,m} \varphi_{\ell m}$ is simply given by $(\varphi_{\ell m})_{,t} = - i \omega_{mn} \varphi_{\ell m}$ (see Eq.~\eqref{psisdec}), while to compute the derivatives with respect to the radial coordinate we note that
\begin{equation}
\begin{cases}
    Y_{\ell m} =  \delta\varphi_{\ell m \omega}^+ e^{i \omega r}&\text{for $r \rightarrow \infty $}\ , \\
Y_{\ell m} =  \delta\varphi_{\ell m \omega}^- e^{-i k r_{\star}}&\text{for $r \rightarrow r_{+}$}\ ,
\end{cases}
\end{equation}
therefore
\begin{equation}
\begin{cases}
   (\varphi^*_{\ell m})_{,r} =  -i \omega_{mn} \varphi^*_{\ell m}  &\text{for $r \rightarrow \infty $}\ ,\\
    (\varphi^*_{\ell m})_{,r} =  i k_{mn} \frac{r^2+a^2}{\Delta} \varphi^*_{\ell m} &\text{for $r \rightarrow r_{+}$}\,.
\end{cases}
\end{equation}
Using these relations and the orthogonality condition of the spheroidal functions \eqref{ortogonality} we obtain the energy fluxes for the scalar field in the frequency domain: 
\begin{align}
    \dot{E}_\tn{scal}^{(+)} &= \frac{1}{16 \pi} \sum_{\ell,m,n} \omega_{mn}^2 |\delta \hat{\varphi}^{+}_{\ell mn}|^2\ , 
    \label{scalfluxinf}\\
    \dot{E}_\tn{scal}^{(-)} &= \frac{1}{16 \pi} \sum_{\ell,m,n} \omega_{mn} k_{mn} |\delta \hat{\varphi}^{-}_{\ell mn}|^2\ , 
    \label{scalfluxhor}
\end{align}
with $k_{mn} \equiv \omega_{mn} - m \Omega_+$. 

From the energy flux we obtain the the angular momentum flux:
\begin{align}
    \dot{L}_\tn{scal}^{(+)} = \frac{1}{16 \pi} \sum_{\ell,m,n} m \omega_{mn} |\delta \hat{\varphi}^{+}_{\ell mn}|^2\ , 
    \label{angularfluxinf}\\
    \dot{L}_\tn{scal}^{(-)} = \frac{1}{16 \pi } \sum_{\ell,m,n} m k_{mn} |\delta \hat{\varphi}^{-}_{\ell mn}|^2\ . 
    \label{angularfluxhor}
\end{align}
The total energy and angular momentum fluxes are given by the sum of the gravitational 
and scalar terms at horizon and at infinity:
\begin{equation}
    \dot{C}_\tn{GW} =  \sum_{i=+,-}[\dot{C}^{(i)}_{\rm grav}+\dot{C}^{(i)}_{\rm scal}]\ =  \dot{C}_\tn{grav} +  \dot{C}_\tn{scal}\ , \label{totflux}
\end{equation}
where $C\in[E,L]$, $\dot{C}_\tn{grav} \equiv \dot{C}_\tn{grav}^{(+)} + \dot{C}_\tn{grav}^{(-)} $ and  $ \dot{C}_\tn{scal} \equiv \dot{C}_\tn{scal}^{(+)} + \dot{C}_\tn{scal}^{(-)}$. 

Finally, because of the linear dependence of the source term 
from the scalar charge, it is worth to remark that 
the total scalar flux for a given scalar charge simply scales 
with $d$ as 
\begin{equation*}
    \dot{C}_\tn{scal} = d^ 2 \dot{\bar{C}}_\tn{scal}\ , 
\end{equation*}
where $\dot{\bar{C}}_\tn{scal}$ only depends on $(p,e,a)$.

\subsection{Adiabatic Orbital Evolution}\label{sec:orbitalevo}
The loss of energy and angular momentum due to the gravitational and the scalar GW emission drives the binary orbital evolution, which in the adiabatic approximation follows a sequence of geodesics until the plunge. 
The change of the orbital integrals $C=(E,L)$ is given by the balance law 
\begin{equation}
   \dot{C} = - \dot{C}_\tn{GW}\ .
\end{equation}
The evolution of $(E,L)$ allows us to study the change of 
the eccentricity and of the semi-latus rectum 
$(e,p)$ \cite{Glampedakis:2002ya}:
\begin{equation}
\dot{E}=E_{,p}\dot{p}+E_{,e}\dot{e}\quad\ ,\quad 
\dot{L}=L_{,p}\dot{p}+L_{,e}\dot{e}\ ,
\end{equation}
such that 
\begin{align}
\dot{p}=(L_{,e}\dot{E}-E_{,e}\dot{L})/H\ \ ,\ \ 
\dot{e}=(E_{,p}\dot{L}-L_{,p}\dot{E})/H\ ,
\end{align}
with $H=E_{,p}L_{,e}-E_{,e}L_{,p}$. 
As shown in Eq.~\eqref{totflux} the scalar flux adds linearly to the gravitational component and increases the rate of change of the orbital parameters $(p,e)$. Therefore, for a given set of initial condition, binaries with $d\neq 0$ complete less cycles before plunge than binaries with vanishing scalar charge.
To quantify the impact of the scalar charge on possible GW detections by LISA, we evolve  EMRIs with and without scalar charge, following their orbits for a given observational time $T_\textnormal{obs}$. 
At any time $t$, we compute the orbital frequencies for both systems, and the  quadrupolar dephasing
\begin{equation}
    \Delta \Psi_{i}= 2 \int^{T_\tn{obs}}_{0} \Delta \Omega_{i} \dd t \qquad i =\phi,r\ , 
\label{eq:deph}
\end{equation}
where 
\begin{equation}
    \Delta \Omega_i = \Omega_i^{d} - \Omega_i^{d=0}. 
\end{equation}

Since the quadrupolar component dominates the dephasing, and $\Delta\Psi_r\ll\Delta\Psi_\phi$, the dephasing of the gravitational wave is $\Delta\Phi\sim\Delta\Psi_\phi$.
Following~\cite{Gupta:2021cno} we choose $\Delta\Psi_{\phi} \sim 0.1\ \text{rad}$ as the threshold for a dephasing observable by LISA for a system detected with SNR of $\sim 30$~\cite{Bonga:2019ycj}.

The dephasing provides a preliminary estimate of the scalar charge distinguishability.
A more quantitive and accurate assessment can be made by computing the faithfulness $\mathcal{F}$ between two GW signals emitted by EMRIs with and without the charge:
\begin{equation}\label{eq:def_F}
\mathcal{F}[h_1,h_2]=\max_{\{t_c,\phi_c\}}\frac{\langle h_1\vert 
	h_2\rangle}{\sqrt{\langle h_1\vert h_1\rangle\langle h_2\vert h_2\rangle}}\ ,
\end{equation}
where we have introduced the noise-weighted inner product between two templates in the frequency domain
\begin{equation}
\langle h_1\vert h_2\rangle=4\Re\int_{f_\tn{min}}^{f_\tn{\rm max}}\frac{\tilde{h}_1(f)\tilde{h}^\star_2(f)}{S_n(f)}\dd f\ , \label{math:sca_prod}
\end{equation}
maximised over time and phase offsets $(t_c,\phi_c)$ between the two signals. We consider the power spectral density $S_n(f)$ of the LISA detector including the confusion noise produced by unresolved galactic white dwarf binaries~\cite{Robson:2018ifk}.
The Fourier transform $\tilde{h}(f)$ of the GW signals is computed starting from the waveform model for eccentric inspirals in the time domain described in~\cite{Barack:2003fp} (see also Appendix~\ref{appendix:waveform} for more details). We also fix the minimum integration frequency to $f_{min}= 10^{-4}$Hz, while $f_{max}=f_\tn{Ny}$, with $f_\tn{Ny}$ being the Nyquist frequency.

Eq.~\eqref{math:sca_prod} also allows to compute the SNR $\rho$ of a signal $h$, $\rho = \langle h\vert h\rangle$. Assuming $\rho = 30$, two signals result to be distinguishable by LISA if $\mathcal{F} \lesssim 0.994$~\cite{Chatziioannou:2017tdw}.

\subsection{Implementation}
\label{numericalintegration}
To compute both the gravitational and the scalar fluxes we have exploited some of the numerical routines implemented in the Black Hole Perturbation Toolkit (BHPT)~\cite{BHPToolkit}, and in particular the \texttt{Teukolsky} package to calculate the homogeneous solutions of the Teukolsky equation. Since BHPT currently assumes circular orbits only, the integration over the source terms needed to obtain the eccentricity-dependent perturbations and the corresponding fluxes at the horizon and infinity has been performed using a Mathematica code built independently. We have checked that for $e=0$ our code reproduces the fluxes obtained by the BHPT with great accuracy. Comparisons with previous results for eccentric EMRIs in GR are discussed in Appendix~\ref{appendix:checks}. Orbital frequencies have also been computed using the BHPT, in particular the  \texttt{KerrGeodesics} package \cite{Fujita:2009bp}.

We have computed $\left(\dot{E}^{(\pm)},\dot{L}^{(\pm)}\right)$ for different values of $(e,p)$ and assuming $a=0.2M$ and $a=0.9M$ for the primary spin. Note that that MBHs falling in the LISA band are expected to be rapidly spinning, with $a\sim 0.9M$ or possibly even larger~\cite{Babak:2017tow}.
We have sampled the eccentricity between $0.1\leq e \leq 0.5$ in steps of $\Delta e = 0.1$. These choices lead to an eccentricity at the plunge in agreement with standard expectation, a flat distribution in $e \in [0,0.2]$~\cite{Babak:2017tow}. To make the grid in $p$ denser close to the separatrix, where orbital parameters vary more rapidly, for the semi-latus we have taken 41 points uniformly spaced in the new variable $u = (p-0.9p_s)^{-1/2}$, within $[u(p_\textnormal{min}),u(p_\textnormal{max})]$, where $p_{max}= p_{min}+10M$ and $p_{min} = p_s + 0.03M$, with $p_s$ being the value of $p$ at the separatrix as a function of $e$. Then, from the inverse relation $p(u)$, we have obtained a non-uniform grid for the semi-latus rectum. In this way, if one considers the grid in $(e,p-p_s)$, the initial and the final values of $p-p_s$ are the same for each values of the eccentricity, i.e. $0.03M$ and $10.03M$, respectively. This is optimal for a two dimensional interpolation with \texttt{Mathematica}, which can be performed only on a structured grid.

For each point in the $(e,p)$ plane we have computed the total flux by summing over the three indexes $(\ell,m,n)$\,\footnote{We remind that the index $n$ is associated to the radial motion, with period $T_r$, see Appendix~\ref{sec:geodesics}.}: 
\begin{equation}
    \dot{C} = \sum_{\ell mn} \dot{C}_{\ell mn} = \sum^{\ell_{max}}_{\ell_{min}} \sum^{m=+\ell}_{m=-\ell} \left(\dot{C}_{\ell m0}+2\sum^{n_{max}}_{n=1} \dot{C}_{\ell mn}\right)\ ,\label{esum}
\end{equation}
where $C\in[E,L]$ and the $\ell=0,1$ components are due to the scalar flux only, while both scalar and gravitational fluxes contribute to the $\ell\ge2$ components.

In our code we have chosen $\ell_\text{max}=(8,10)$ respectively for $s=(0,-2)$. These values are such that, for a primary spin of $a=0.9M$ and eccentricity $e=0.5$, the relative difference in the flux between $\ell_{max}$ and $\ell_{max}-1$ is less than $2\%$ for the innermost $p$ of the grid, while for the outermost is less than $0.01\%$. For $e=0.1$ the relative difference is less than $1\%$ for the innermost $p$ and less than $0.001\%$ for the outermost.

The value for $n_{max}$ is chosen such that the fractional change in the sum \eqref{esum} is smaller than $10^{-4}$ for three consecutive values of $n$. This choice is motivated by the behavior of the energy flux spectrum as a function of $(e,n)$. Indeed, we observe that for low eccentricities the flux has a peak at small values of $n$, rapidly decreasing afterwards. On the other hand, for larger values of $e$ the spectrum shows relative maxima before reaching the absolute peak, located at higher $n$ compared to the low-eccentricity case. This behaviour is shown in Fig.~\ref{fig:spettro} of Appendix \ref{appendix:checks}, where we plot $\dot{E}^{(+)}_{\ell mn}$ as a function of $n$ for different values of the eccentricity and for the $\ell=m=2$ and $\ell=m=5$ modes.

After computing the fluxes for each point of the grid, we have performed an interpolation using 
a built-in \texttt{Mathematica} function over the grid in the two parameters $(e, p-p_s)$. In order to estimate the errors introduced by the interpolation, we computed also the fluxes in points outside the grid, and estimated the relative difference between the interpolated and the computed fluxes. The relative difference between them turns out to be $ \lesssim 0.2\%$ for points fluxes closer to the separatrix and it grows for larger values of $p$, up to $\sim 6-7 \%$ for the furthermost points.
Numerical values for the fluxes obtained for different points outside our grid are listed 
in Tables \ref{tab:interpolationE} and \ref{tab:interpolationL} of Appendix \ref{appendix:interpolation}.

Finally, regarding the waveform templates, we have employed the quadrupolar formula discussed in~\cite{Barack:2003fp} summing over the harmonics with $\ell=2$ and different values of $n$:
\begin{equation}
    h(t) = \sum_{n=1}^{\bar{n}} h_{n}(t)\ .
\end{equation}
Each term is a sum over $m=-2,\dots,2$. We consider contributions up to $\bar{n}=10$. We have checked that the relative difference between the faithfulness computed with this setup, and including a further component, i.e. with $\bar{n}=11$ is $\ll 0.1\%$.

\section{Results}
\label{sec:results}
We shall now discuss how the scalar charge affects the EMRIs orbital 
evolution. We first focus on the case of circular orbits with a 
spinning primary. 

\subsection{Circular orbits}

\begin{figure}[htbp!]
\includegraphics[scale = 0.34]{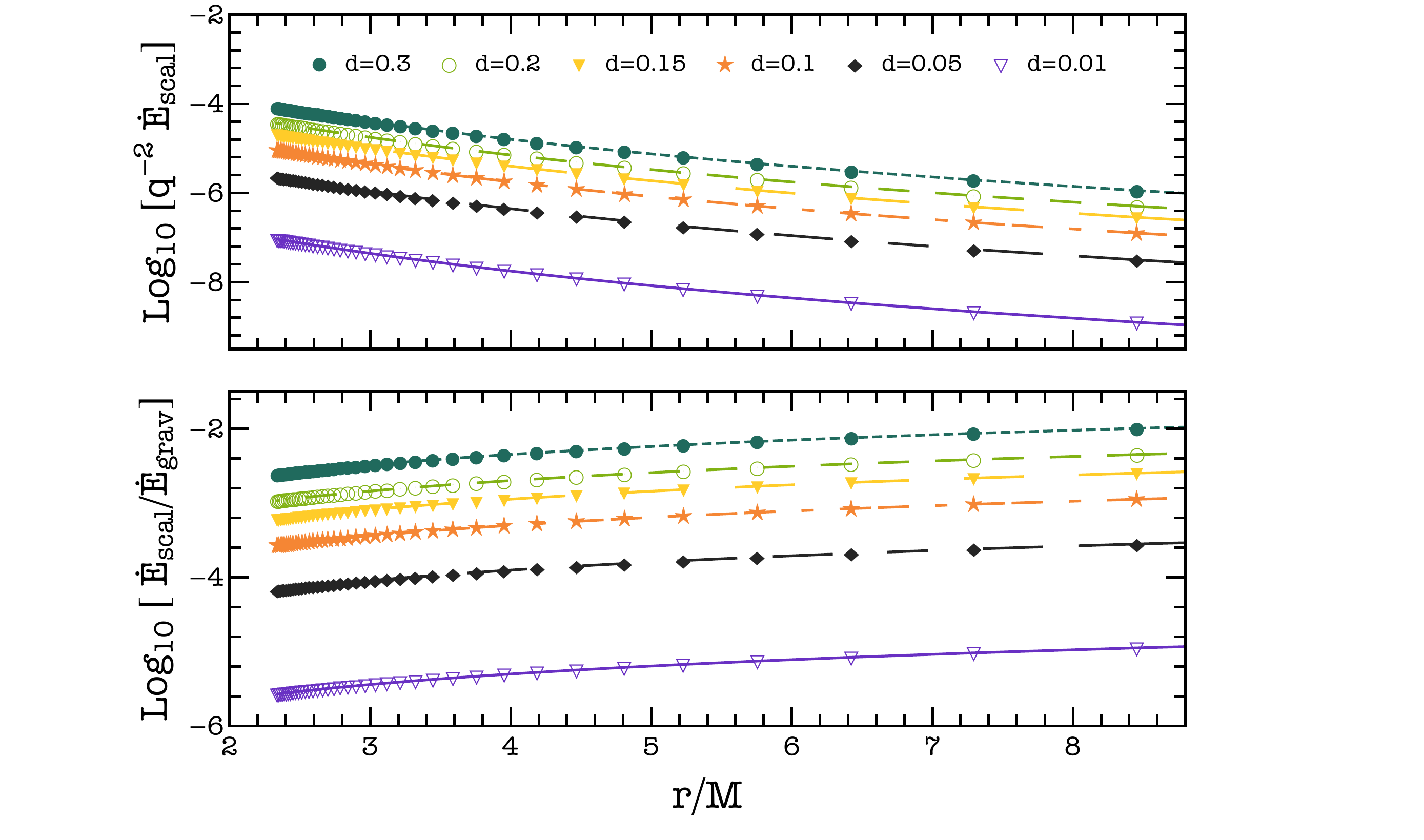}
\caption{(Top) Scalar energy flux, normalized with the mass-ratio, as a function of the orbital radius, for different values of the scalar charge. The spin of the primary is $a/M=0.9$. (Bottom) Relative difference between the scalar and gravitational energy flux as a function of the orbital radius for different values of the scalar charge, and $a/M=0.9$.} 
\label{fig:fluxcirc1}
\end{figure}

\begin{figure}[htbp!]
\includegraphics[scale = 0.42]{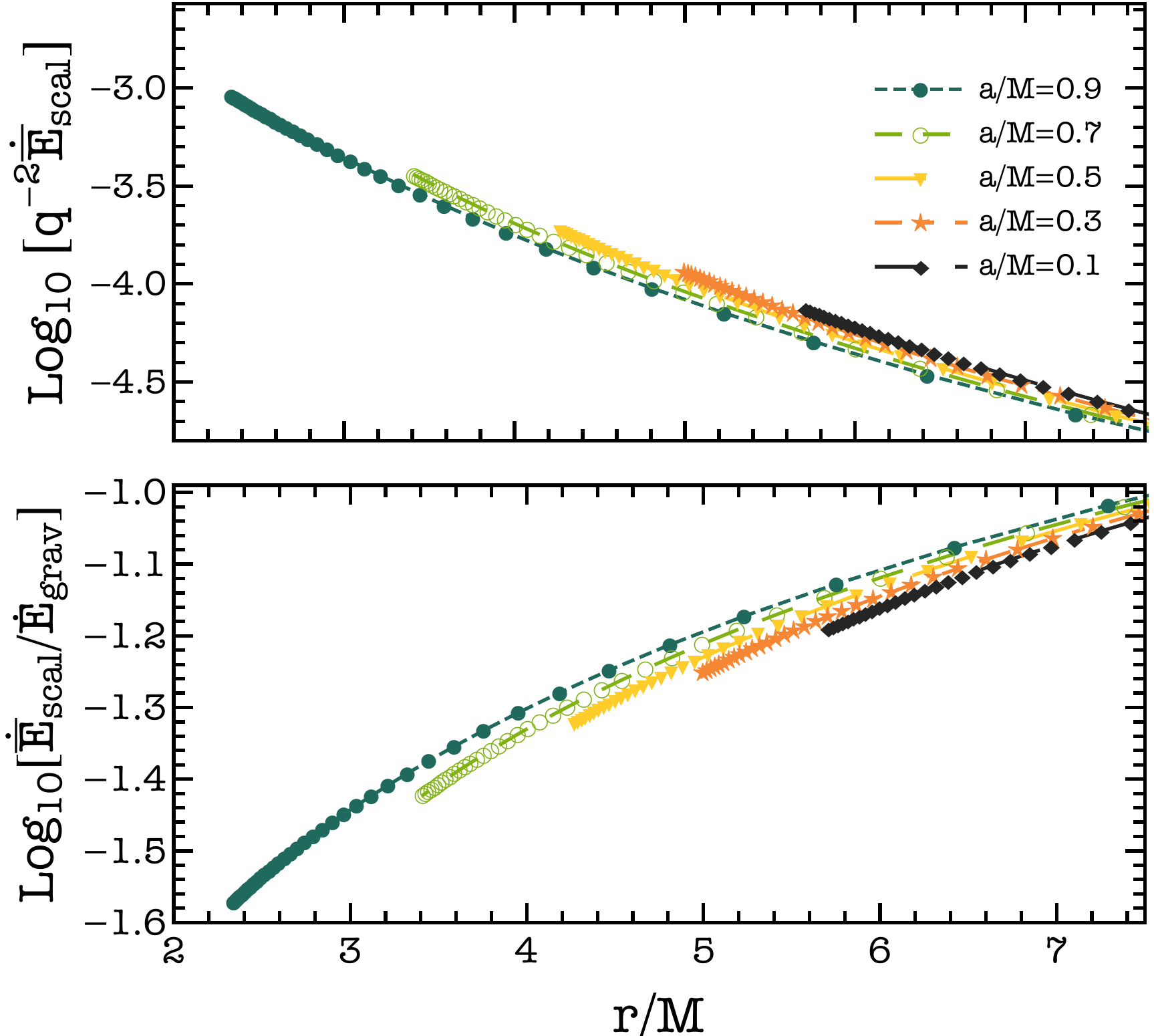}
\caption{Same as Fig.~\ref{fig:fluxcirc1} for different values of the primary spin $a/M$. } \label{fig:fluxcirc2}
\end{figure}

Results for circular orbits are summarized in Figs.~\ref{fig:fluxcirc1}-\ref{fig:fluxcirc2}. The top panel of Fig.~\ref{fig:fluxcirc1} shows the total scalar energy flux as a function of the orbital radius, for different values of $d$ and for primary's angular momentum $a/M=0.9$. Both the scalar and gravitational fluxes scale as $q^2$.
The behavior of $\dot{E}_\tn{scal}$ is qualitatively similar to that shown in paper I, where the central BH is non-spinning. 

The bottom panel of Fig.~\ref{fig:fluxcirc1} shows the ratio between the scalar and gravitational components of the GW flux, for the same binary configurations of the top picture. The ratio decreases as the orbital radius shrinks, with the gravitational contribution growing in time faster than the scalar contribution, at small separation.

Figure~\ref{fig:fluxcirc1} also shows that for $d> 0.01$  the scalar flux $\dot{E}_\tn{scal}$ ranges between $0.1\%$ and $1\%$ of the gravitational flux $\dot{E}_\tn{grav}$. Therefore, we expect the scalar charge to induce a significant contribution on the EMRI evolution when integrated over the all inspiral phase (see paper I).

 Figure~\ref{fig:fluxcirc2} shows the ratio $\dot{E}_\tn{scal}/\dot{E}_\tn{grav}$, rescaled by $d^2$, as a function of the primary BH spin, and the absolute value of $\dot{E}_\tn{scal}$. It is interesting to note that while for a fixed radius $r/M$, larger $a/M$ lead to slightly smaller values of the scalar flux (this is also true for the gravitational component), the overall emission increases due to the larger range of frequencies spanned by the binary. 

\begin{figure}[htbp!]
    \centering
    \includegraphics[scale = 0.4]{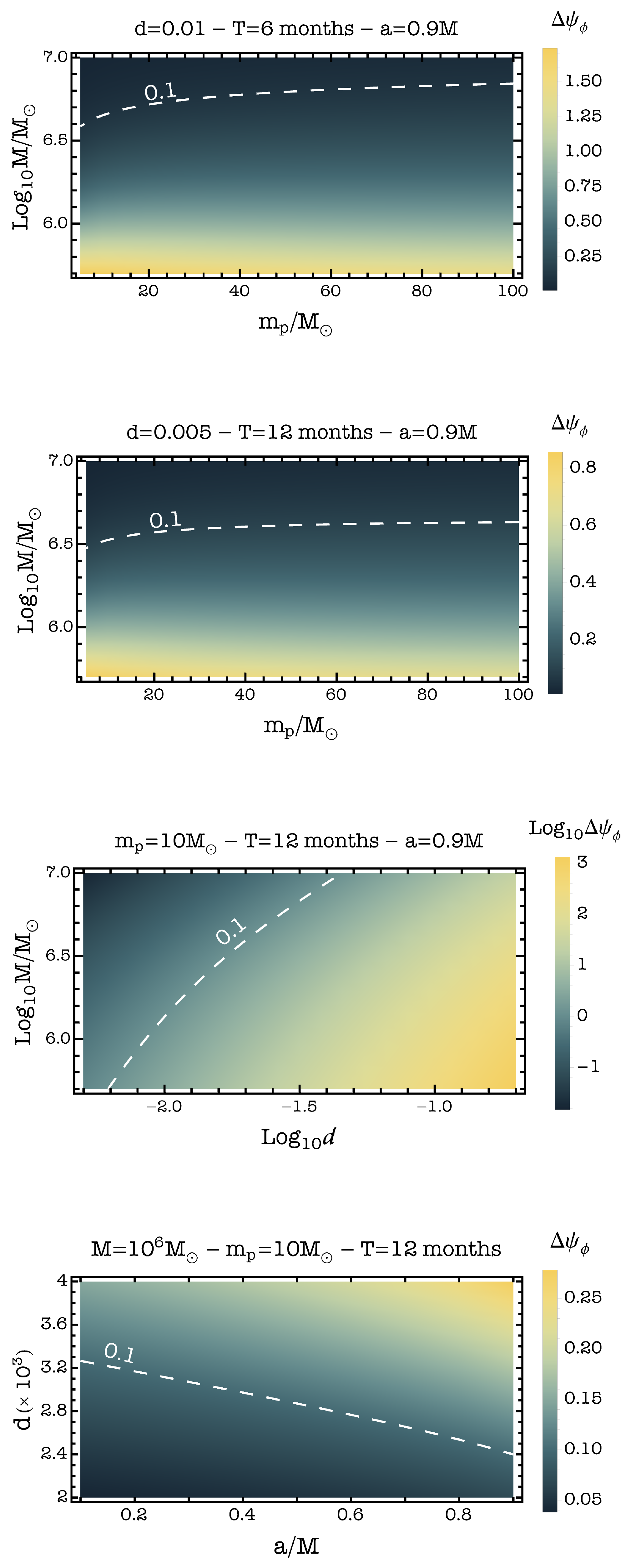}
    \caption{Quadrupolar gravitational wave dephasing $\Delta\Psi_{\phi}$, i.e  difference in the GW phase evolution of EMRIs with and without scalar charge. 
    First and second panels show $\Delta\Psi_{\phi}$ as a function of the binary component masses and refer to EMRIs with $(d,T_\tn{obs})=(0.01,6\ \tn{months})$ and $(d,T_\tn{obs})=(0.005,12\ \tn{months})$, respectively, for $a/M=0.9$. 
    Third and fourth panels show the dephasing as a function of $(M,d)$ and of $(d,a/M)$, respectively, for $T_\tn{obs}=12$ months of observation and $m_\tn{p}=10M_\odot$.
    The dashed white line in each plot identifies the detectability threshold of $0.1$ radian for a GW event with SNR of 30 observed by LISA.}
    \label{fig:circ_deph}
\end{figure}

Figure~\ref{fig:circ_deph} provides the GW dephasing $\Delta\Psi_{\phi}$ defined in Eq.~\eqref{deltaphi}. In the top two panels the dephasing is shown for different values of the scalar charge and of the observation time before the plunge,\footnote{We
define the plunge as $r_\tn{plunge}=r_\tn{ISCO}+\delta r$, where we adopt the conservative choice of $\delta r=0.1 M$ (see paper II). The initial radius $r_0$ for the evolution of each binary is chosen such that the system reaches $r_\tn{plunge}$ from $r_0$ after $T_\tn{obs}.$} $(d,T_\tn{obs})=(0.01,6\ \tn{months})$ and $(d,T_\tn{obs})=(0.005,12\ \tn{months})$,
as a function of  the binary component masses.
All binaries with $M\lesssim 3\times10^6M_\odot$ lead to a dephasing larger than the detectability threshold of $0.1$ radians, with the values of the dephasing $\Delta \Psi_\phi$ being almost insensitive to the mass of the secondary. In both cases, with further six months of observation time (i.e., $T_\tn{obs}=6$ months for $d=0.01$ and $T_\tn{obs}=12$ months for $d=0.005$) all the binary configurations up to $M\simeq10^7\,M_\odot$ and $m_p\simeq100\,M_\odot$ are above the threshold.

In the third panel from the top we study how the dephasing changes as a function of the scalar charge and of the mass of the primary, for $m_p=10M_\odot$, $T_{\tn{obs}}=12$ months and $a/M=0.9$. The plot shows that the accumulated phase difference can be significant, especially for binaries with a massive BH of $M\lesssim 10^6M_\odot$ for which $\Delta\Psi_\phi$ can be larger than $10^3$ radians. 

Finally, the last (bottom) panel of Figure~\ref{fig:circ_deph} shows how $\Delta\Psi_\phi$ changes by varying the spin of the primary and the scalar charge of the secondary. The masses of the binary are $(M, m_p)=(10^{6},10)M_{\odot}$ and the time of observation is $T_{\tn{obs}}=12$ months. For a fixed scalar charge, the dephasing increases with the increasing of the primary spin. For $a/M =0.1(0.9)$, $\Delta \Psi_{\phi}$ is larger then the threshold of $0.1$ radians for $d \gtrsim 0.0033(0.0023)$, respectively. This result is consistent with those of Ref.~\cite{Guo:2022euk}, where the increase of the dephasing with the spin of the primary was discussed.

Overall this analysis confirms the calculations of Papers I and II, and is extremely encouraging in terms of the future constraints on the charge that can be inferred by LISA. 

\subsection{Eccentric orbits}
\begin{figure*}[htbp!]
    \centering
    \includegraphics[scale = 0.4]{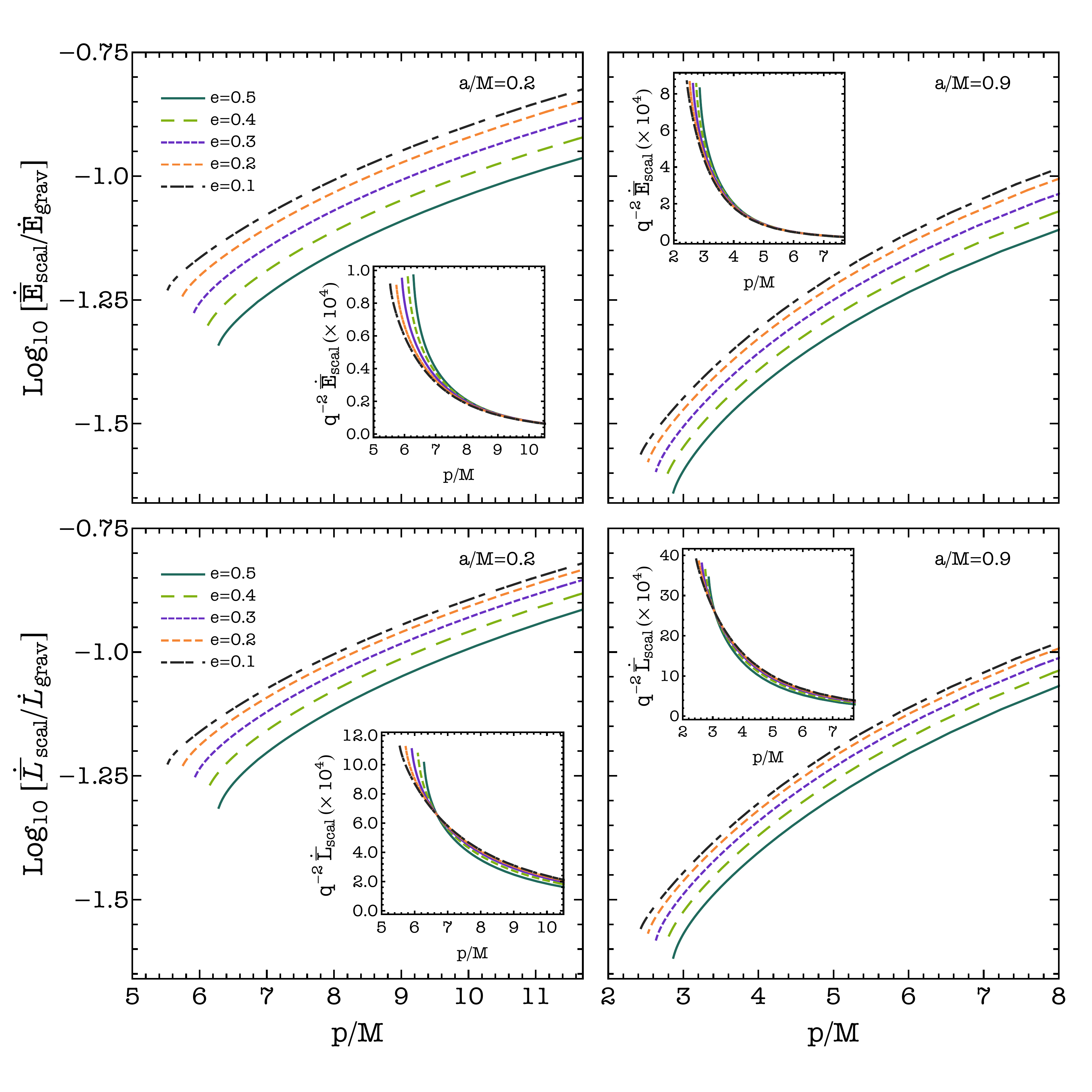}
    \caption{Ratio between the scalar and gravitational energy (top panels) and angular momentum (bottom panels) fluxes as a function of the semi-latus rectum $p$, for different values of the eccentricity and a fixed spin of $a=0.2M$ (left panels) and $a=0.9M$ (right panels). 
    The inset within each plot shows the absolute value of the scalar component, normalized with the mass ratio. We assume $d=1$ for all the configurations.} 
    \label{fig:scalar_ecc_fluxes}
\end{figure*}

We now move to eccentric orbits. 
In Fig.~\ref{fig:scalar_ecc_fluxes} we show the ratio between the scalar and gravitational energy (top panels) and angular momentum (bottom panels) fluxes, as a function of the semi-latus rectum $p/M$, during the inspiral of EMRIs on eccentric orbits with a scalar charge $d=1$.
The inset in each panel provides the absolute value of $\dot{E}_\tn{scal}$ and $\dot{L}_\tn{scal}$. We focus here on two prototype binaries with primary spin $a/M=0.2$ (left panels) and $a/M=0.9$ (right panels), and eccentricities $e\in[0.1,0.5]$.

\begin{figure}[htbp!]
    \centering
    \includegraphics[scale = 0.34]{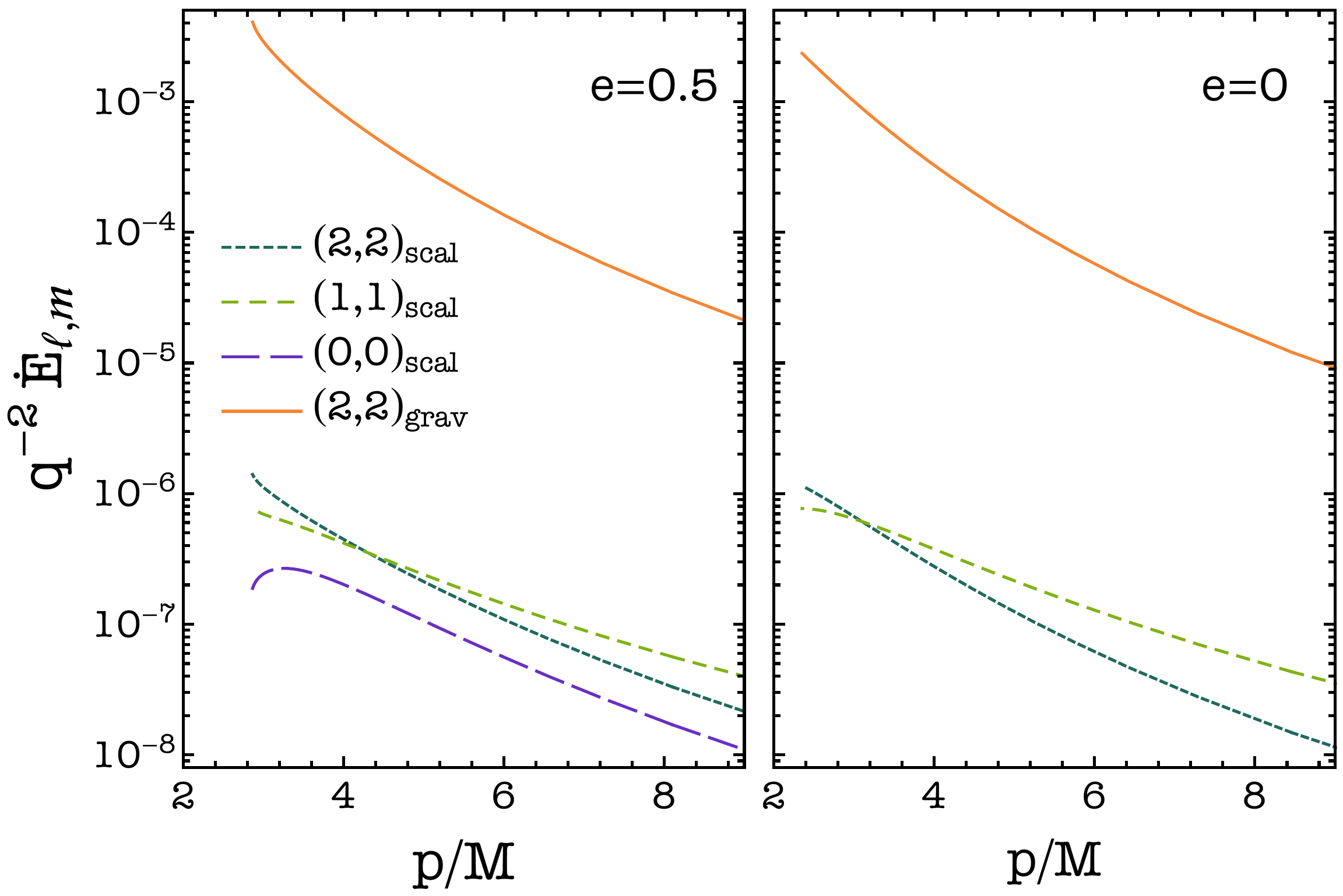}
    \caption{Harmonic components of the gravitational and scalar energy flux, normalized with the mass-ratio, with $d=0.1$ as a function of the semi-latus rectum, for eccentric (left panel) and circular (right panel) orbits. The MBH's spin is $a/M = 0.9$.
    We only show the dominant modes, i.e. the $\ell=(0,1,2)$ coefficients and, for the eccentric orbit, we sum over the index $n$. In the gravitational sector only $\ell \geq2$ modes are present, while in the scalar sector the dipole contribution is excited. The monopole contribution of the scalar sector is only excited in the eccentric case. 
    \label{fig:multipoles_e05}}
\end{figure}

As was the case for circular orbits discussed above, for a given value of the eccentricity the ratio between the scalar and the gravitational components decreases for smaller $p$, due to the faster growth of $\dot{E}_\tn{grav}$ and $\dot{L}_\tn{grav}$. This behavior is also  confirmed by the analyses of the harmonic components shown in Fig.~\ref{fig:multipoles_e05} for $e\neq 0$. 
Moreover, for fixed $p$, while the absolute value of $\dot{E}_\tn{scal}$ grows with the eccentricity, the relative difference with respect to the gravitational flux becomes smaller. Note that the value of the separatrix increases for higher eccentricity. However, the periastron of the last stable orbit decreases for higher eccentricity, so that a more eccentric orbit can lead the particle closer to the MBH horizon.

As shown in Fig.~\ref{fig:multipoles_e05}, for large orbital separation, the dipole $\ell=m=1$ scalar mode approaches the quadrupolar $\ell=m=2$ (scalar) 
component, with the latter increasing steeply for 
smaller separations. The monopole coomponent $\ell=m=0$ is excited only for eccentric orbits and shows a similar steep increase, although it remains subdominant and starts decreasing before the plunge.

\begin{figure}[htbp!]
    \centering
    \includegraphics[scale = 0.35]{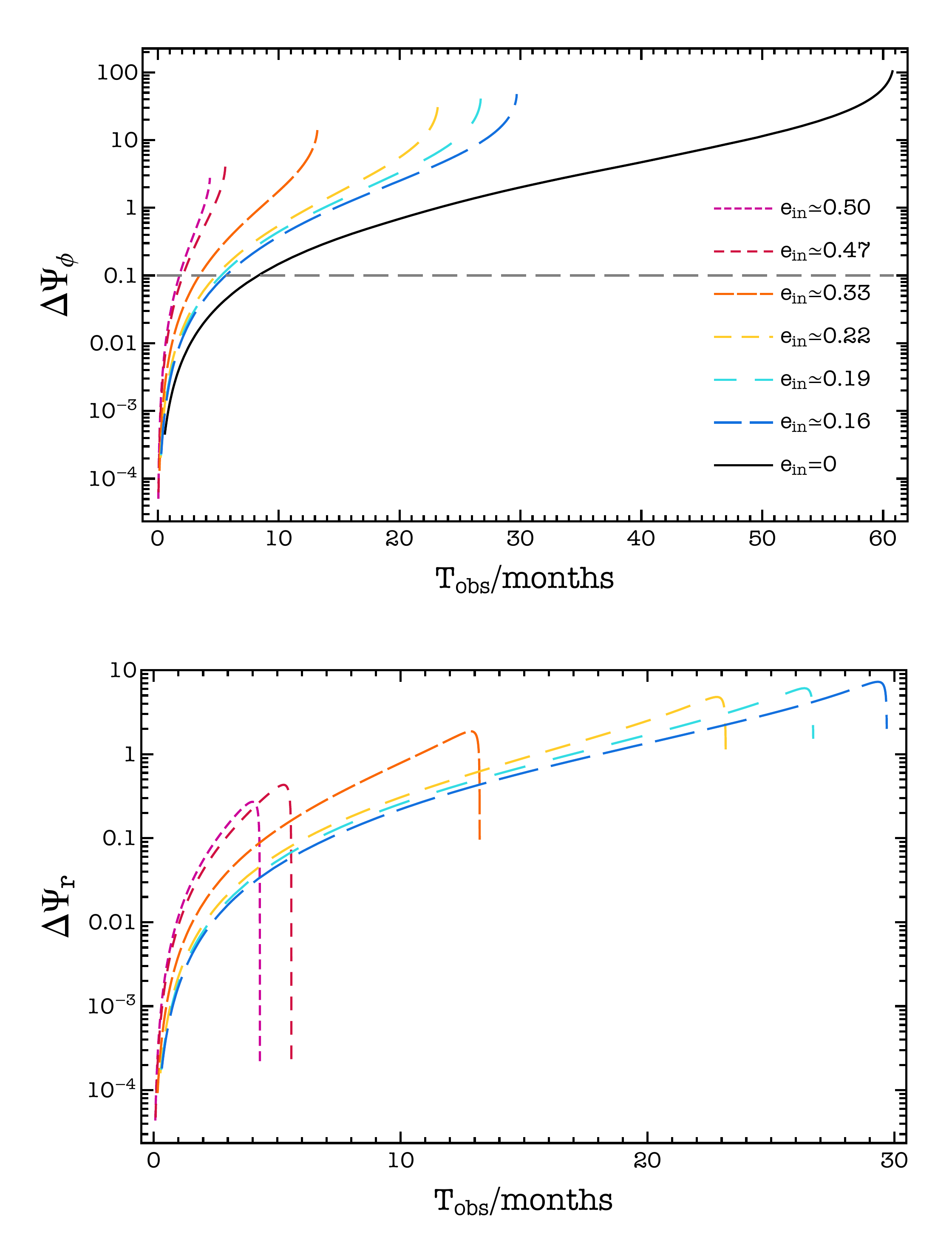}
    \caption{Azimuthal (top) and radial (bottom) quadrupolar dephasing as a function of the observation time, assuming spin $a=0.9M$. The curves refer to inspirals with initial apastron $r_a=11M$ and different values of the initial periastron, which correspond to initial eccentricities within $e\simeq[0,0.5]$.
    We fix the scalar charge to $d=0.01$. The horizontal line in the top panel identifies the threshold for phase resolution by LISA for a binary observed with signal to SNR of 30~\cite{Bonga:2019ycj}. Although this is not visible in the semi-logarithmic scale, the first two curves in the bottom panel reach negative values.
 }
    \label{fig:scalar_ecc_deph}
\end{figure}

In order to quantify the impact of the eccentricity on the EMRI evolution we analyse, for various orbital set ups, the quadrupolar dephasing induced by $d$. The latter is computed by comparing two different inspirals, with and without the scalar charge, starting with the same initial condition, {\em i.e.}~with the same initial periastron and apastron, and assuming initial phases $\Psi_\phi=\Psi_r=0$. The phases are defined by
\begin{equation}
    \Psi_i=\int_0^{T_{\rm obs}}\Omega_idt~~~(i=\phi,r)\, .
    \label{eq:single_phases}
\end{equation}
The values of $\Delta\Psi_{\phi,r}$ are shown in Fig.~\ref{fig:scalar_ecc_deph} for the total orbital evolution up to the plunge\footnote{Here we mean the position $(e_{fin}, p_{fin})$ such that $p_{fin}=p_{min}(e_{fin})$, with $p_{min}=p_s+0.11M$.}. The dashed curves refer to binary configurations with fixed apastron $r_a=11 M$, and different choices of the periastron $r_p$. The scalar charge has been fixed to $d=0.01$. The initial and final values of the periastron and eccentricity of each inspiral with $d=0.01$ are provided in Table \ref{tab:data_inspiral}. 
We observe that, by increasing the initial periastron, 
i.e. by reducing the initial eccentricity, 
the time it takes for the secondary to reach 
the plunge grows, leading to larger values of the 
accumulated dephasing. However, for a given time of observation, $\Delta \Psi_\phi$ is larger for 
inspirals with higher $e_{in}$.

In order to assess the detectability by LISA, we plot the $\Delta \Psi_\phi =0.1$\,rad threshold, corresponding to the minimum phase potentially resolvable by the detector for a binary observed with a SNR of $30$ (note that $\Delta\Psi_r\ll\Delta\Psi_\phi$ and thus gives a negligible contribution to the dephasing).
After $4$-$6$ months of observation all the considered inspirals lead to a dephasing larger then the threshold. Values of the scalar charge $>0.01$ will also lead to larger dephasing. We provide some reference values of the latter after $12$ months of evolution for different choices of $d$ in Table~\ref{tab:data}. 
The steep variation in $\Delta \Psi_r$ at the end of the evolution appears to be due the 
orbital eccentricity, whose time derivative 
changes signs close to the plunge. That eccentricity grows as one approaches the plunge has already been pointed out in the literature, see {\em e.g.}~\cite{Glampedakis:2002ya}. While $e(t)$ increases, 
the change of $\Psi_r$ (defined in Eq.~\eqref{eq:single_phases}) for 
$d\neq 0$ becomes smaller then the one for $d=0$, 
$\Delta\Psi_r$ acquires a negative sign and ends up counterbalancing the 
dephasing accumulated until the turning-point.

\begin{figure}[htbp!]
    \centering
    \includegraphics[scale = 0.5]{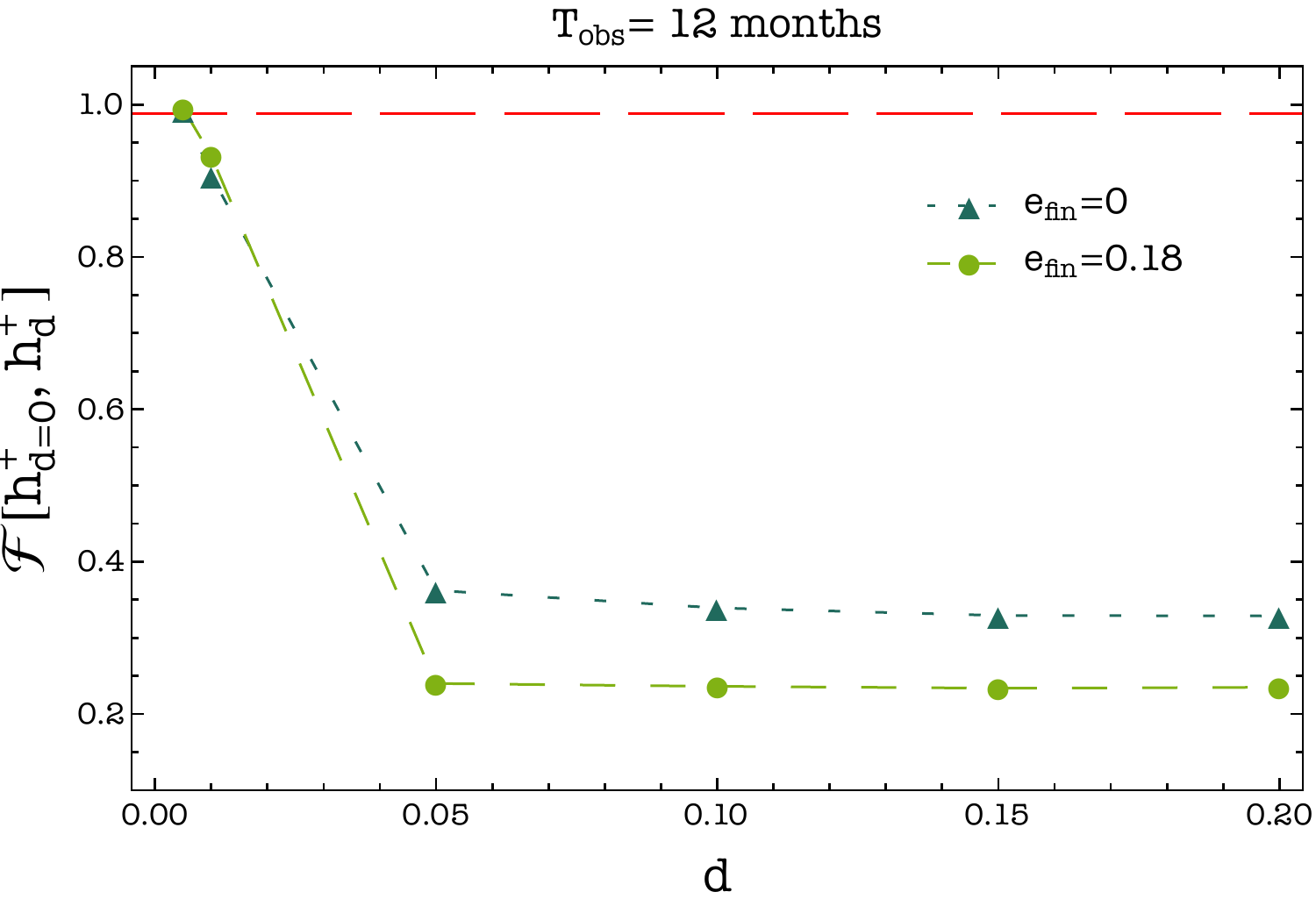}
    \caption{Faithfulness between the plus polarization of 
    two GW waveforms computed with $d=0$ and $d \neq 0$, as a function of the scalar charge, for a circular (dark triangles) and an eccentric (light dots) inspiral. The MBH's spin is $a/M = 0.9$, while the time of observation is fixed to 12 months. The dashed line corresponds to the threshold below which the two templates are distinguished by LISA for a binary observed with SNR of 30~\cite{Chatziioannou:2017tdw}.}
    \label{fig:faithfulness}
\end{figure}

\begin{table}[htbp!]
\begin{tabular}{cccc}
\hline
\hline
 $r^{in}_p/M$  & $e_{in}$ &  $r^{fin}_p/M$ &  $e_{fin}$ \\
\hline
 3.667 & 0.49997  & 2.18606  & 0.18945\\ 
 4 & 0.47  & 2.19251 & 0.18185\\
 5.5  & 0.33  &2.19721& 0.17846 \\
 7 & 0.22  & 2.19756 & 0.17812\\ 
 7.5 & 0.19  & 2.19759& 0.17809\\
 7.9  & 0.16  & 2.19761 & 0.17807\\
 11  & 0  & 2.789 & 0\\
  \hline
  \hline
\end{tabular}
\caption{Values of the initial and final periastron and eccentricity for the inspirals of Fig.~\ref{fig:scalar_ecc_deph} }\label{tab:data_inspiral}
\end{table}

\begin{table}[htbp!]
\begin{tabular}{ccccc}
\hline
\hline
 $e_{in}$ & $d$  & $\Delta \Psi_\phi$ & $\Delta \Psi_r$ \\
\hline
0.22 & 0.01  & 0.88   & 0.5\\ 
     & 0.05  & 21  & 12\\
     & 0.1   & 88 & 48 \\   
0.33 & 0.01  & 4   & 1.5\\ 
     & 0.05  & 105  & 38\\
     & 0.1   & 423 & 151 \\
  \hline
  \hline
\end{tabular}
\caption{Values of the accumulated dephasings after $12$ months 
of evolution for three different values of the scalar charge 
$d = (0.01,0.05,0.1)$, for a primary spin $a/M=0.9$. The initial apastron is fixed to $r_a=11M$, as 
for the plots in Fig.~\ref{fig:scalar_ecc_deph}.
}\label{tab:data}
\end{table}

Finally, we compute the faithfulness between the plus polarization of two GW templates with $d=0$ and $d\neq 0$, assuming $12$ months of observation. The faithfulness is shown in Fig.~\ref{fig:faithfulness}, as a function of the scalar charge $d$, comparing a circular and an eccentric inspiral around a MBH of spin $a/M=0.9$.  The initial position for each inspiral is found by requiring that the secondary reaches the plunge after one year of evolution, fixing the final eccentricity to $e_{fin}=0.18$ such that the 
plunge is located at $(p_{fin}/M,e_{fin})\simeq(2.59, 0.18)$. In particular, for $d=0$ the initial position is given by $(p_{in}/M, e_{in})\simeq(7.071,0.492)$. For the sake of 
comparison we also compute the faithfulness in the case of circular orbits, 
for which the plunge is fixed at the ISCO.

The horizontal dashed line in the figure represents the threshold value, beyond which signals with SNR of $30$ can be distinguished. While for $d \gtrsim 0.01$ the faithfulness for both the circular and the eccentric inspiral is well below the threshold, we see that for the latter the distinguishability increases, leading to a smaller overlap between the the templates. 

While a more sophisticated analysis is required to determine the actual constraints on $d$ that can be inferred by EMRI on eccentric orbits (see paper II), our dephasing and faithfulness results provide a strong indication  that LISA should be able to constrain or detect even small values of the  scalar charge. Moreover,  Fig.~\ref{fig:faithfulness} suggests that  the inclusion of the eccentricity in the analysis improves the distinguishability of the scalar charge, at least for $d\gtrsim0.01$.

\section{Conclusions}\label{sec:conclusions}

EMRIs are golden sources for the future GW space-based detector LISA. Due to their rich phenomenology, they are ideal for  investigating a large variety of astrophysical phenomena and for testing fundamental physics. 
Probing the behavior of gravity in a strong-field dynamical regime is a major science goal for LISA, which will benefit from the observation of the hundreds of thousands of GW cycles that the EMRI secondary will follow before plunging into the massive central body.

Tracking the long orbital inspiral of EMRIs is a complex task which requires accurate templates to be compared against actual data~\cite{Pound:2015tma,Barack:2018yvs,Pound:2019lzj,Warburton:2021kwk}. 
Until recently, EMRI GW templates have only been developed in GR. Such templates only allow for null tests of GR: performing unbiased tests of gravity requires the development of templates which include the effects of possible deviations from GR or the Standard Model~\cite{Yunes:2009ke}. On the other hand, beyond-GR templates would in general be theory-dependent, and may be extremely challenging to compute within the required accuracy.

In paper I we have shown that these problems can be overcome for a large class of theories with an additional massless scalar field. Indeed, at leading order in the mass ratio, the MBH spacetime in these theories is described by the Kerr metric, and the changes in the EMRI dynamics due to the presence of the scalar field only depend on the scalar charge of the inspiralling body, which uniquely captures all the information on the underlying theory of gravity.  
In paper II we have included in our analysis the spin of the primary body and we have assessed, using a Fisher-matrix approach, the capability of LISA to measure the scalar charge.
Here we have generalized some of the results of these papers, by studying in detail the effect of rotation on the signal, and by including the eccentricity of the (prograde) orbits in the model.

The spin of the primary has a strong influence on the GW emission of the binary. Indeed, increasing the MBH rotation rate, the radius at which the secondary plunges shrinks, allowing it to inspiral closer to massive BH and then enhancing the overall energy loss. We have shown that for a fixed orbital distance the ratio between the scalar and gravitational fluxes also increases with the spin: for $a/M=0.9$ the value of $\dot{E}_\tn{scal}/d^2$ can be as large as $\sim10\%$ of $\dot{E}_\tn{grav}$ far from the plunge ($r/M\sim 8$), decreasing to $\sim2\%$ for closer distances. 
The difference in the phase evolution between EMRIs modelled with and without scalar charge suggests that LISA can be potentially able to identify values of $d$ as small as $ d \sim 0.0033$, for $a/M=0.1$ and $d \sim 0.0023$, for $a/M=0.9$, for one year of observations before the plunge. Large values of the charge may lead to significant dephasing of more than $10^3$ radians for the same observing time.

We have also studied the EMRI evolution on eccentric inspirals, computing the scalar emission for various orbital configurations, and the induced GW dephasing.
Comparing different inspirals with the same initial apastron our results show that for a given time of observation the phase difference increases for larger values of the initial eccentricity.

Furthermore, we have investigated the distinguishability of GW signals emitted by eccentric EMRIs carrying a non-vanishing scalar charge, by computing the faithfulness between waveforms with different values of $d$. We confirm the small values of the faithfulness previously found in paper II, finding that the eccentricity further reduces the overlap with respect to the equatorial circular case. Our analysis suggests that one year of observation by LISA would be enough to distinguish signals with a scalar charge as small as $d\simeq0.01$. Hence, the joint effect of the eccentricity and of the MBH spin, both expected for EMRIs in real astrophysical environments, enhance the dephasing with respect to the uncharged case, and leads to promising results in terms of LISA observations. As already discussed in paper II, the dephasing only provides a preliminary assessment of the detectability of the scalar charge. The faithfulness analysis is rather more robust, but a more complete and accurate analysis which take into account the correlation between all binary parameters is needed. Studies based on MCMC simulations and on state of the art EMRI waveform generation~\cite{Katz:2021yft} are underway.

We remark that our encouraging results are, qualitatively, not limited 
to the class of theories considered in this paper.
Indeed, if the MBH is not described by the Kerr metric, the deviations are expected to  be even larger, and our results can be considered as a conservative estimate. 

The template developed here already provides a significant improvement of those derived in papers I and II, but a more refined analysis is still needed in order  to assess LISA’s full potential to detect fundamental fields and new physics beyond GR. Besides taking into account statistical correlations as discussed above, our analysis should be extended to retrograde or inclined orbits, exploiting previous self-force calculations for scalar 
charges on generic trajectories \cite{Warburton:2014bya, Nasipak:2019hxh, Drasco:2005kz}, to the case of an (ultralight) massive scalar field (see~\cite{papermassive}), and to theories with multiple fields and couplings constants.
A more challenging task is given by the extension of our formalism
beyond the adiabatic order, where the coupling between the 
gravitational and the scalar field will introduce new 
computational 
problems, potentially leading to significant 
changes to the EMRI orbital evolution.


\noindent{{\bf{\em Acknowledgments.}}}
A.M.~and T.P.S.~thank Niels Warburton for useful 
discussions, and for kindly sharing previous results 
on numerical computations of the 
scalar flux. This work makes use of the Black Hole Perturbation Toolkit.
The authors would like to acknowledge networking 
support by the COST Action CA16104. A.M.~acknowledge 
support from the Amaldi Research Center funded by 
the MIUR program "Dipartimento di Eccellenza" 
(CUP: B81I18001170001). N.F.~acknowledges financial support provided under the European Union's H2020 ERC Consolidator Grant ``GRavity from Astrophysical to Microscopic Scales'' grant agreement no.~GRAMS-815673. We also acknowledge financial fupport from the EU Horizon 2020 Research and Innovation Programme under the Marie Sklodowska-Curie Grant Agreement no.~101007855. T.P.S.~acknowledges partial support from the STFC Consolidated Grant no.~ST/T000732/1 and no.~ST/V005596/1.

\appendix

\section{Geodesic motion}\label{sec:geodesics}
We model the EMRI dynamics as the motion of a test body 
moving along geodesics of a Kerr BH with intrinsic angular momentum $J=aM$. In the Boyer-Lindquist 
coordinates $x^{\mu} = (t,r,\theta,\phi)$, the Kerr metric reads 
\begin{multline}
    \dd s^2 = - \left(1 - \frac{2Mr}{\Sigma}\right)\dd t^2 - \frac{4aMr \sin^2\!{\theta}}{\Sigma} \dd t \dd\phi + \frac{\Sigma}{\Delta} \dd r^2 \\+ \Sigma \dd\theta^2 + \left(r^2 + a^2 + \frac{2Ma^2r\sin^2{\theta}}{\Sigma}\right)\sin^2\!{\theta}\,\dd\phi^2,
\end{multline}
where $\Delta (r ) \equiv r^2 - 2Mr + a^2 $ and $\Sigma(r, \theta) \equiv r^2 + a^2 \cos^2{\theta}$~\cite{Visser:2007fj}. The outer event horizon of the 
BH is located at $r_{+} = M + \sqrt{M^2 - a^2}$.

In this paper we consider equatorial eccentric orbits, such that 
the geodesic equations are given by: 
 \begin{align}
     r^2 \frac{\dd r}{\dd\tau}&=\pm (V_r)^{1/2}= \pm \sqrt{T^2 - \Delta \left[r^2+(L-aE)^2\right]}\ ,\\
     r^2 \frac{\dd\phi}{\dd\tau}&= -(aE-L)+\frac{aT}{\Delta}\ ,\\
        r^2 \frac{\dd t}{\dd\tau}&= -a (aE-L)+ \frac{\left(r^2+a^2\right)T}{\Delta}\ ,\\
        \theta(\tau)&=\pi/2\ , 
 \end{align} 
where $\tau$ is the proper time of the secondary and 
$T \equiv E\left(r^2+a^2\right)-a L $~\cite{Hughes:1999bq}. The space-time admits 
two constants of motion, $E$ and $L$, which correspond to the energy 
and the angular momentum of the particle at infinity. Once initial 
condition are specified, $(E,L)$ uniquely determine a 
bound equatorial orbit in the Kerr space-time. Orbits are 
defined by $0\leq E < 1$ and confined between the periastron 
$r_p$ and the apastron $r_a$, being $r_p \leq r \leq r_a$, which 
represent the turning points of the orbital motion, and such that 
$V_r(r_a)=V(r_p)=0$. Equatorial orbits can be parametrised either 
by the energy and angular momentum, or by two parameters $p$ and $e$, 
i.e., by the semi-latus rectum and by the eccentricity of the orbit, 
with $0\leq e < 1$. These parameters are defined in terms of the 
turning points as
\begin{equation}
    r_p = \frac{p}{1+e}, \qquad r_a = \frac{p}{1-e}\ . 
\end{equation}
The relation between $(p,e)$ and $(E,L)$, as well as an explicit  
expression for the latter are shown later in this Appendix. 
The radial coordinate can also be parametrised in terms of 
a new parameter $\chi$: 
\begin{equation}
    r(\chi) = \frac{p}{1+e\cos{\chi}}\ , 
\end{equation}  
where $\chi$ varies monotonically from $\chi=0$ at the periastron, 
to $\chi=\pi$ at the apastron. 
We define the radial period $T_r$ as the coordinate time taken from 
the secondary to pass through two consecutive periastron passages, 
such that $T_r = t(\chi=2 \pi)=2t(\chi=\pi)$. 
Given $T_r$ the period of the radial motion, we also introduce 
$\Delta \phi$ as the variation of $\phi$ in an interval $T_r$. 
The functions $t(r)$ and $\phi(r)$ can be obtained by integrating 
the geodesics equations: 
\begin{align}
    t(r) &= \int^{r}_{r1}\frac{\dd t}{\dd\tau} \left(\frac{\dd r}{\dd\tau}\right)^{-1}  \dd r\ , \\
    \phi(r)&=\int^r_{r1}\frac{\dd\phi}{\dd\tau}\left(\frac{\dd r}{\dd\tau}\right)^{-1}  \dd r\ .
\end{align}
However, these integrals result to be divergent at the 
turning points of the orbit. To avoid this divergence, we 
can perform the integrals over the parameter $\chi$: 
\begin{align}
\phi(\chi)&= \int^\chi_0 \dd\chi' \frac{\tilde{V}_\phi(\chi',p,e)}{ J(\chi',p,e)\tilde{V}_r(\chi',p,e)}\ ,
\label{eq:phichi}\\
t(\chi)&= \int^\chi_0 \dd\chi' \frac{\tilde{V}_t(\chi',p,e)}{ J(\chi',p,e)\tilde{V}_r(\chi',p,e)}\ ,
\label{eq:tchi}
\end{align}
where the functions $\hat{V}_{t,r,\phi}$, amd $J$  are defined as 
\begin{align}
    \tilde{V}_r&=x^2 + a^2 + 2 a x E - \frac{2 M x^2}{p}(3+e \cos{\chi})\ ,\\
    \tilde{V}_\phi&= x + aE - \frac{2 M x}{p}(1+e \cos{\chi})\ ,\\
    \tilde{V}_t&= a^2 E - \frac{2 a M x}{p}(1+e \cos{\chi})+\frac{E p^2}{(1 + e \cos{\chi})^2}\ , \\
    J&= 1 - \frac{2 M}{p}(1 + e \cos{\chi})+\frac{a^2}{p^2}(1+e\cos{\chi})^2\ ,
\end{align}
with $x\equiv L-aE$. 

From the radial period $T_r$ and the variation of $\phi$ in such period, we define the orbital frequencies $\Omega_r$ and $\Omega_\phi$ as 
\begin{equation}
    \Omega_r = \frac{2\pi}{T_r}, \qquad   \Omega_\phi = \frac{\Delta \phi}{T_r}\ .
    \label{eq:frequencies_def}
\end{equation}

As we will show, the phase of the emitted gravitational wave signal 
will be related to $(\Omega_r,\Omega_\phi)$ through 
the frequency $\omega_{mn}$:  
\begin{equation}
    \omega_{mn} = m \Omega_\phi + n \Omega_r\ , 
    \label{eq:omegamn}
\end{equation}
with $(m,n)\in\mathbb{Z}$. 
In this work we considered only prograde 
orbits\footnote{In the case of circular orbits $(e=0)$, 
the GW phase depends on 
\begin{equation}
\omega_m = m \omega_p\ ,
\end{equation}
with $\omega_p $ being the angular velocity of the particle
\begin{equation}
    \omega_{p} = \frac{d\phi}{dt} = \pm \frac{M^{1/2}}{r^{3/2}\pm a M^{1/2}}\ , 
    \label{omegakerr}
\end{equation} 
where the $+$ ($-$) sign holds for the prograde (retorgade) 
orbits.}. The frequencies have been computed by making use of the BHPT~\cite{Fujita:2009bp}.
 
A typical inspiral will proceed as a sequence of eccentric 
geodesics, progressively closer to the primary, until the 
radial coordinate reaches the Last Stable Bound Orbit (LSBO) 
beyond which the secondary will plunge~\cite{Ori:2000zn}. 
All bound equatorial orbits have $p^2 > x^2 (1+e)(3-e)$, 
where $x=x(a,p,e)$. Given a certain value of the spin $a$ 
and of the eccentricity $e$, the curve $p^2_s = x^2 (1+e)(3-e)$ 
defines the separatrix in the $e-p$ plane. 
If $e=0$, the separatrix reduces to the ISCO in the Kerr 
spacetime
\begin{equation}
    r_\textnormal{ISCO}/M = 3+Z_2\pm\sqrt{\left(3-Z_1\right)\left(3+Z_1+2Z_2\right)}\ ,
\end{equation}
where the terms $Z_1$ and $Z_2$ are given by the expressions 
$Z_1 = 1+\sqrt[3]{1-(a/M)^2}\left(\sqrt[3]{1+(a/M)}+\sqrt[3]{1-(a/M)}\right)$ 
and $Z_2 = \sqrt{3(a/M)^2+Z_1^2}$~\cite{Bardeen:1972fi}. For $a/M=0$ we obtain the Schwarzschild 
limit $r_\textnormal{ISCO}=6M$. 
For a given value of the spin $a$, the energy fluxes emitted by the binary are computed until the secondary reaches $r_\textnormal{ISCO}$ 
or the separatrix, depending on whether we 
assume circular or eccentric orbits.

The expressions for the energy $E$ and angular momentum $L$ as a function of $(p,e)$ are given by: 

\begin{equation}
    E = \left[ 1 - \left(\frac{M}{p}\right) (1-e^2) \left(1-\frac{x^2}{p^2}(1-e^2)\right)\right]^{1/2}\ ,
\end{equation}

\begin{equation}
   L = x + a E \ ,
\end{equation}
where
\begin{equation}
 x = \left[\frac{-N-\tn{sign}(a)\sqrt{N^2-4FC}}{2F}\right]^{1/2}\ ,
\end{equation}
and the functions $F,N,C$ are given by:
\begin{align}
 F(p,e) =& \frac{1}{p^3}\big[p^3-2M(3+e^2)p^2+M^2(3+e^2)^2p\nonumber\\
 &-4Ma^2(1-e^2)^2\big]\ ,\\
 N(p,e) =&\frac{2}{p}\big\{\left[M^2(3+e^2)-a^2\right]p-Mp^2\nonumber\\
 &-Ma^2(1+3e^2)\big\}\ ,\\
 C(p) =& (a^2-Mp)^2\ .
\end{align}

\section{Gravitational perturbations}
\label{appendix:gravitational_perturbations}

As discussed in Sec.~\ref{sec:assumptions} the field equations 
for the gravitational field are the same of those for a Kerr 
BH in GR. Therefore, in computing the gravitational 
perturbations we follow the approach pioneered by Teukolsky~\cite{Teukolsky:1973ha}, and the work described in~\cite{Glampedakis:2002ya}. For $s=-2$ Eqns.~\eqref{teukeq}, 
\eqref{psisdec} and \eqref{sourcedec} 
lead to a differential equation for $R_{lm\omega} \equiv R_{lm}(\omega,r)$: 
\begin{equation}
    \Delta^2 \frac{\dd}{\dd r} \left( \frac{1}{\Delta} \frac{\dd R_{lm\omega}}{\dd r}\right) - V(r)R_{lm\omega}(r) = J_{lm\omega}(r)\ ,
    \label{eq}
\end{equation}
where the potential is given by 
\begin{equation}
    V(r)= - \frac{K^2 + 4 i (r-M)K}{\Delta}+8i\omega r + \lambda\ , 
    \label{potkerr}
    \end{equation} 
with $K= (r^2+a^2)\omega - ma$ and $\lambda$ is the angular eigenvalue of Eq. \eqref{eq:spherharm}. The source term $J_{lm\omega}$ is described in Sec.~\ref{sourceterm_teuk}. 
As done for the scalar case we define $Y = \Delta^{-1} \sqrt{r^2+a^2} R$ 
and we find that the homogeneous solution $Y_-$, which satisfies the 
condition of purely ingoing wave at the horizon, and $Y_{+}$, 
which satisfies the condition of purely outgoing wave at infinity, 
have the following asymptotic behaviour: 
\begin{equation}
\begin{cases}
    Y_- = B^{hole}_{lm\omega}e^{-ikr_\star} &\text{for $r \rightarrow r_{+}$}\ ,\\
    Y_- = \frac{B^{in}_{lm\omega}}{r^2} e^{-i\omega r_\star} + B^{out}_{lm\omega}r^2 e^{i\omega r_\star} &\text{for $r \rightarrow \infty$}\ ,
\end{cases}
\end{equation}

\begin{equation}
\begin{cases}
    Y_+ = D^{in}_{lm\omega} e^{-ikr_\star}+ D^{out}_{lm\omega} e^{i k r_\star} &\text{for $r \rightarrow r_{+}$}\ ,\\
    Y_+ = D^{\infty}_{lm\omega}r^2 e^{i\omega r_\star} &\text{for $r \rightarrow \infty$}\ ,
\end{cases}
\end{equation}
where $k = \omega - m \omega_+$ and $\omega_+ = a/(2Mr_+)$. 
From the relation \eqref{y} between $R$ and $Y$ we 
obtain the boundary condition for the homogeneous solutions: 

\begin{equation}
\begin{cases}
    R^{-}_{lm\omega} = B^{hole}_{lm\omega} \Delta^2 e^{-i k r^{*}} &\text{for $r \rightarrow r_{+}$}\ ,\\
    R^{-}_{lm\omega} =  \frac{B^{in}_{lm\omega}}{r} e^{-i\omega r^{*}}+B^{out}_{lm\omega} r^3 e^{i \omega r^{*}} &\text{for $r \rightarrow \infty$}\ ,\label{rhout}
    \end{cases}
\end{equation}

\begin{equation}
    \begin{cases}
    R^{+}_{lm\omega} =  D^{in}_{lm\omega} \Delta^2 e^{- i k r^{*}}+D^{out}_{lm\omega} e^{i k r^{*}}&\text{for $r \rightarrow r_{+}$}\ ,\\
    R^{+}_{lm\omega} = D^{\infty}_{lm\omega} r^3 e^{i \omega r^{*}}&\text{for $r \rightarrow \infty$}\ .
    \end{cases}
\end{equation}

Finally, the general solution for $R_{lm\omega}$ is given by 
$R_{lm\omega}(r) = Z^-_{lm\omega}(r)R^{+}_{lm\omega}(r)+ Z^{+}_{lm\omega}(r)R^{-}_{lm\omega}(r)$ where, following~\cite{Hughes:1999bq},
\begin{align}
Z^-_{lm\omega}(r) &= \frac{1}{2 i \omega B^{in}_{lm\omega} D^{\infty}_{lm\omega}} \int_{r_+}^{r}{ \dd r' \frac{R^-_{lm\omega}(r')I_{lm\omega}(r')}{\Delta(r')^2}}\ ,\\
Z^+_{lm\omega}(r) &= \frac{1}{2 i \omega B^{in}_{lm\omega} D^{\infty}_{lm\omega}} \int_{r}^{\infty}{ \dd r' \frac{R^+_{lm\omega}(r')I_{lm\omega}(r')}{\Delta(r')^2}}\ .
\end{align}
Defining $Z^{-}_{lm\omega} \equiv Z^{-}_{lm\omega}(r \rightarrow \infty)$, $Z^{+}_{lm\omega} \equiv Z^{+}_{lm\omega}(r\rightarrow r_+)$, 
the asymptotic radial solutions read: 
\begin{align}
    R_{lm\omega}(r \rightarrow \infty) &= Z^{-}_{lm\omega} D^{\infty}_{lm\omega} r^3 e^{i \omega r^{*}}\ ,\\
    R_{lm\omega}(r \rightarrow r_+) &= Z^{+}_{lm\omega} B^{hole}_{lm\omega} \Delta^2 e^{-i k r^{*}}\ . 
\end{align}
It is convenient to absorb the factors $D^{\infty}_{lm\omega}$ and 
$B^{hole}_{lm\omega}$ into $Z^{-}_{lm\omega}$ and $Z^{+}_{lm\omega}$, 
such that: 

\begin{align}
Z^-_{lm\omega}(r) &= \frac{1}{2 i \omega B^{in}_{lm\omega}} \int_{r_+}^{r}{ \dd r' \frac{R^-_{lm\omega}(r')I_{lm\omega}(r')}{\Delta(r')^2}}\label{zh}\ ,\\
Z^+_{lm\omega}(r) &= \frac{B^{hole}_{lm\omega}}{2 i \omega B^{in}_{lm\omega} D^{\infty}_{lm\omega}} \int_{r}^{\infty}{ \dd r' \frac{R^+_{lm\omega}(r')I_{lm\omega}(r')}{\Delta(r')^2}}\ ,\label{zinf}
\end{align}
and 
\begin{align}
    R_{lm\omega}(r \rightarrow \infty) &= Z^{-}_{lm\omega} r^3 e^{i \omega r^{*}}\ ,
    \label{rinf}\\
    R_{lm\omega}(r \rightarrow r_+) &= Z^{+}_{lm\omega} \Delta^2 e^{-i k r^{*}}\ . 
\end{align}
The coefficients $Z^{-,+}_{lm\omega}$ are needed to 
calculate the energy flux at horizon and at infinity. 
However, since in Eq. \eqref{rhout} the outgoing solution grows 
with a coefficient $r^4$ relative to the ingoing coefficient, the 
ingoing solution is completely swamped, and obtaining 
$B^{in}_{lm}$ is extremely challenging. The reason for this 
difficulty is that the potential $V(r)$ in Eq.~\eqref{potkerr} 
of the Teukolsky equation is long ranged. The solution for 
this problem relies in transforming 
the Teukolsky equation into the Sasaki-Nakamura equation~\cite{Sasaki:1981sx}, which features a short-ranged potential. Other possible methods to solve the Teukolsky equation are the Mano-Suzuki-Takasugi method \cite{Sasaki:2003xr}, which is the one used in the BHPT, and an approach that makes use of a hyperboloidal foliation \cite{Zenginoglu:2011jz, Piovano:2021iwv}.

\subsection{Energy Flux}

The energy flux for gravitational waves can be computed in terms of the Isaacson stress-energy tensor~\cite{Isaacson:1968zza}, and is given by:

\begin{equation}
    \left(\frac{\dd^2 E }{\dd A \dd t}\right)^\text{rad}_{r \rightarrow \infty} = \frac{1}{16 \pi} \bigg \langle \left( \frac{\partial h_+}{\partial t}\right)^2 + \left(\frac{\partial h_{\times}}{\partial t}\right)^2 \bigg \rangle \ ,
    \label{isaaaaaaac}
\end{equation}
where the brackets $\big \langle ... \big \rangle$ denote an 
average over a region of spacetime large compered with the 
wavelenght of the radiation. The expression for $h_+$ and 
$h_\times$ are obtained considering that, for $r\rightarrow \infty$, 
the value of $\Psi_4$ is given by: 
\begin{equation}
    \Psi_4 (r \rightarrow \infty ) = \frac{1}{2}\left( \ddot{h}_+ - i \ddot{h}_{\times}\right)\ .
\end{equation}
To obtain the gravitational energy flux we integrate 
\eqref{isaaaaaaac} with respect to time variable. 
The expression for $\Psi_4$ is obtained combining 
Eq.~\eqref{psipsi4}- \eqref{psisdec}, with $R_{lm\omega}$ 
given by \eqref{rinf} in the limit $r\rightarrow \infty$. 
Combining all these equations we obtain the energy flux at infinity as: 
\begin{align}
    \left( \frac{dE}{dt}\right)^{rad}_{r \rightarrow \infty} &= \sum_{lmn}{\frac{|Z^-_{lmn}|^2}{4 \pi \omega_{mn}^2}}\ ,\\
      \left( \frac{dL}{dt}\right)^{rad}_{r \rightarrow \infty} &= \sum_{lmn}{\frac{m |Z^-_{lmn}|^2}{4 \pi \omega_{mn}^3}}
    \label{fluxinfkerr}
\end{align}

Where $\omega_{mn}$ is given in \eqref{eq:omegamn} and the explicit form of $Z^{H}_{lmn}$ is given in the next 
section. 
The energy flux at the horizon can be calculated by measuring the rate at which the event horizon's area increases as radiation falls into it, following the prescription of~\cite{hawking1972} as described in~\cite{Teukolsky:1973ha}. The result, given by~\cite{Hughes:1999bq} 
reads: 
\begin{align}
    \left( \frac{dE}{dt}\right)^{rad}_{r \rightarrow r_+} = \sum_{lmn}{\alpha_{lmn}\frac{|Z^{+}_{lmn}|^2}{4 \pi \omega_{mn}^2}}\ ,\\
    \left( \frac{dL}{dt}\right)^{rad}_{r \rightarrow r_+} = \sum_{lmn}{\alpha_{lmn}\frac{m |Z^{+}_{lmn}|^2}{4 \pi \omega_{mn}^3}}\,
    \label{fluxhorkerr}
\end{align}
where the coefficients $\alpha_{lmn}$ are given by
\begin{equation}
   \alpha_{lmn} = \frac{256 (2Mr_+)^5 k_{mn} (k_{mn}^2+4\epsilon^2)(k_{mn}^2+16\epsilon^2)\omega_{mn}^3}{|C_{lmn}|^2}\ ,
\end{equation}
with $k_{mn} = \omega_{mn} - m \omega_+$, 
\begin{equation}
    \epsilon = \frac{\sqrt{M^2-a^2}}{4Mr_+}\ ,
\end{equation}
and
\begin{align}
|C_{lmn}|^2 = &[(\lambda+2)^2+4a\omega_{mn}-4a^2\omega_{mn}^2](\lambda^2\nonumber\\
&+36ma\omega_{mn}-36a^2\omega_{mn}^2) \nonumber\\
&+(2\lambda+3)(96a^2\omega_{mn}^2-48ma\omega_{mn})\nonumber\\
&+144\omega_{mn}^2(M^2-a^2)\ . 
\end{align}

\subsection{The source term}
\label{sourceterm_teuk}

In this section we show here the explicit expression for the 
source terms needed compute the amplitudes 
$Z^{-,+}_{\ell m \omega}$  for eccentric orbital 
configurations. We refer the reader to~\cite{Glampedakis:2002ya} 
for more details on thir derivation.
The amplitudes $Z^{-,+}_{\ell m\omega}$ \eqref{zh}-\eqref{zinf} 
can be obtained computing 

\begin{equation}
Z^{-,+}_{\ell m\omega} = \frac{m_p}{2 i \omega B^{in}}\int^{\infty}_{-\infty} \dd t e^{i \omega t - i m \phi(t)} I^{-,+}_{\ell m\omega}[r(t),\theta(t)]\ , 
\label{eq:ilm}
\end{equation}

where

\begin{align}
    I^{-,+}_{\ell m\omega} =& \left[ R^{in,up}_{lm\omega} \{A_{nn0} + A_{\bar{m}n0}+A_{\bar{m}\bar{m}0}\} -  \right. \nonumber \\  & \left. - \frac{dR^{in,up}_{lm\omega}}{dr}  \{A_{\bar{m}n1}+A_{\bar{m}\bar{m}1}\}+ \right. \nonumber\\
    & \left. + \frac{d^2 R^{in,up}_{lm\omega}}{dr^2}A_{\bar{m}\bar{m}2} \right]_{r=r(t),\theta = \theta(t)}\ . 
\end{align}

The expression for the coefficients $A$ are given by 

\begin{widetext}

\begin{align}
    A_{\bar{m}n0}(u) =& \frac{2}{\sqrt{\pi}}\frac{C_{\bar{m}n}}{u(1-2Mu+a^2u^2)^2}\left[2a^2u^3+\{ia(a\omega-m)-4M\}u^2+2u+i\omega\right] \times \nonumber \\
    &\times \left[\frac{\partial S}{\partial \theta}(\pi/2)+(a\omega-m)S_{lm}(\pi/2)\right]\ , \\
    A_{\bar{m}\bar{m}0}(u) =& \frac{1}{\sqrt{2\pi}}\frac{C_{\bar{m}\bar{m}}S(\pi/2)}{u^2(1-2Mu +a^2u^2)^2}\bigg\{-2ia^3 (a\omega-m)u^5+a(a\omega-m)\{6iM+a(a\omega-m)\}u^4 \nonumber\\
    &-4ia(a\omega-m)u^3+2\omega\{iM+a(a\omega-m)\}u^2-2i\omega u +\omega^2\bigg\}\ , \\
    A_{\bar{m}n1}(u) =& \frac{2}{\sqrt{\pi}}\frac{C_{\bar{m}n}}{u(1-2Mu+a^2u^2)}\left[\frac{\partial S}{\partial \theta}(\pi/2)+(a\omega-m)S(\pi/2)\right]\ , \\
    A_{\bar{m}\bar{m}1}(u) =& -\sqrt{\frac{2}{\pi}}\frac{C_{\bar{m}\bar{m}}S(\pi/2)}{u^2 (1-2Mu +a^2u^2)}\left[a^2u^3+\{ia(a\omega-m)-2M\}u^2+u+i\omega\right]\ ,\\
    A_{\bar{m}\bar{m}2}(u) =& -\frac{1}{\sqrt{2\pi}}\frac{C_{\bar{m}\bar{m}}S(\pi/2)}{u^2}\ ,\\
    A_{nn0}(u) =& -\sqrt{\frac{2}{\pi}}\frac{C_{nn}}{(1-2Mu+a^2u^2)^2}\Bigg\{-2ia \left(\frac{\partial S}{\partial \theta}(\pi/2)+(a\omega-m)S(\pi/2)\right)u   \nonumber\\  &+\frac{\partial^2S}{\partial \theta^2}(\pi/2)+2(a\omega-m)\frac{\partial S}{\partial \theta}(\pi/2)+\{(a\omega-m)^2-2\}S(\pi/2) \Bigg\}\ ,
\end{align}

where $u(\chi,p,e) = (1+e \cos{\chi})/p$ and 
 
\begin{align*}
    C_{nn}(\chi,p,e) &=  \frac{J(\chi,p,e)}{4 p^4 V_{t}(\chi,p,e)}\left[p^2 E -ax (1+e\cos{\chi})^2+ep\sin{\chi}\sqrt{V_r(\chi,p,e)}\right]^2 ,\\
    C_{\bar{m}n}(\chi,p,e) &=  \frac{i x J(\chi,p,e)}{2 \sqrt{2} p^3 V_{t}(\chi,p,e)}(1+e\cos{\chi})\left[p^2 E -ax (1+e\cos{\chi})^2+ep\sin{\chi}\sqrt{V_r(\chi,p,e)}\right]\ ,\\
    C_{\bar{m}\bar{m}}(\chi,p,e) &= - \frac{x^2 J(\chi,p,e)}{2 p^2 V_{t}(\chi,p,e)}(1+e\cos{\chi})^2\ . 
\end{align*}

Finally, we note that recasting Eq.~\eqref{eq:ilm} in term of the variable $\chi$, 
the integral can be written as 

\begin{equation}
    Z^{+,-}_{lmk} = \frac{m_p \Omega_r}{2 i \omega_{mn}B^{in}}\int^{\pi}_{0} \dd\chi \frac{V_t(\chi)}{J(\chi)\sqrt{V_r(\chi)}}\left[I^{+,-}_{lm\omega (+)}(r(\chi))e^{i\omega_{mn}t(\chi)-im\phi(\chi)}+ I^{+,-}_{lm\omega(-)}(r(\chi))e^{-i\omega_{mn}t(\chi)+im\phi(\chi)}\right]\ , 
\end{equation}

where the subscripts $(\pm)$ imply the substitution $\sin{\chi}\rightarrow \pm \sin{\chi}$ in the functions $I^{+,-}_{\ell m \omega}$. 
\end{widetext}

\section{Comparison with previous results}
\label{appendix:checks}

We have tested the numerical output of our code 
by comparing the energy and angular momentum fluxes 
for the scalar and gravitational sector against know 
results published in 
literature \cite{Gralla:2005et, Warburton:2010eq, Warburton:2011hp}. 
In Table~\ref{tab:confsch}
we provide a comparison showing the relative 
difference $\delta$ between the 
calculations of $\dot{E}_\tn{scal,grav}$ and of 
$\dot{L}_\tn{scal,grav}$ for different EMRI orbital set up.
Moreover, in Table~\ref{tab:conf_multipoles} we also show 
the comparison of our results for the scalar emission on eccentric orbits with $a/M=0$, $p=8M$ and $e=0.1$, with
an independent computation (courtesy of N. Warburton computed 
using the code of \cite{Warburton:2011hp}).
For all configurations considered, our results provide a 
remarkable agreement with previous computations.

\begin{figure}
\includegraphics[width=1\columnwidth]{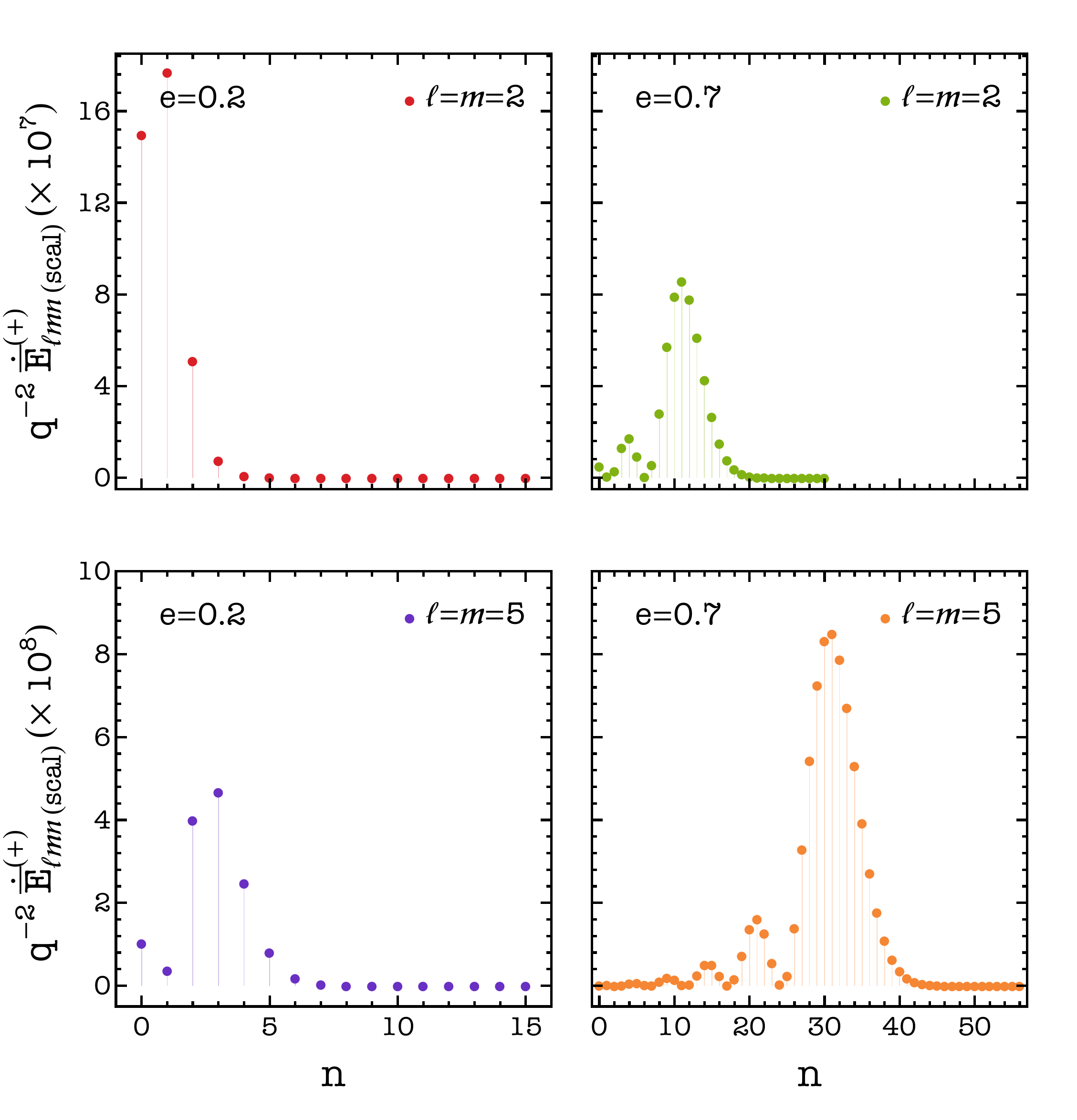}
\caption{Harmonic components of the scalar energy flux at infinity as a function of $n$, for orbital configurations with $p/M=7$, $e=0.2$ (left column) and $e=0.7$ (right column). Top and bottom panels show the $\ell=m=2$ and $\ell=m=5$ components of the flux, respectively. The MBH spin is $a/M=0.9$.
}. \label{fig:spettro}
\end{figure}

Finally, in Fig.~\ref{fig:spettro} we show the behaviour of the energy flux with the index $n$ by plotting the components 
$\dot{\bar{E}}^{(+)}_{\ell m n(\tn{scal})}$ as a function of $n$ for different values of the eccentricity and for the $\ell=m=2$ and $\ell=m=5$ modes. Note that as $\ell=m$ increases, the peak also appears at larger values of $n$. The 
flux components at the horizon shows a similar behavior.

\begin{widetext}

\begin{table}[htbp!]
\begin{tabular}{ccccccccc}
\hline
\hline
reference & sector & $a$ & $p$  & $e$ & $\dot{E}$ & $\delta$\% & $\dot{L} $ & $\delta$\%\\
\hline
\cite{Warburton:2011hp} & scal - (tot) & 0.9  & 10  & 0.2 & 2.686e-5 &   $\simeq3\ee$-$5\%$ &  8.359e-4 &   $\simeq3\ee$-$5\%$\\
\cite{Warburton:2011hp} & scal - (tot) & 0.9  & 10  & 0.5 & 2.468e-5 & $\simeq7\ee$-$1\%$& 6.296e-4&   $\simeq3\ee$-$4\%$  \\
\cite{Warburton:2011hp} & scal - (tot) & 0 & 10 & 0.2  & 3.213e-5 & $\simeq4\ee$-$5\%$ &  9.626e-4 & $\simeq3\ee$-$5\%$\\
\cite{Warburton:2011hp} & scal - (tot) & 0 & 10 & 0.5 & 3.329e-5 & $\simeq1\ee$-$3\%$ & 7.845e-4&$\simeq6\ee$-$4\%$ \\
\cite{Warburton:2011hp} & scal - (tot) & 0.2  & 6.15& 0.4  & 3.427e-4  & $\simeq3\ee$-$2\%$ & 3.926e-3 &  $\simeq2\ee$-$2\%$\\
\cite{Glampedakis:2002ya} & grav - $(+)$ & 0.9  & 12.152  & 0.3731 &  2.737e-5 &$\simeq14\%$\\
\cite{Glampedakis:2002ya} & grav - $(+)$ & 0.5 & 6 & 0.1 & 7.106e-4 &  $\simeq2\ee$-$3\% $  & 1.055e-2 & $\simeq2\ee$-$3\% $ \\
\cite{Glampedakis:2002ya} & grav - $(+)$ & 0.5 & 6 & 0.2 & 7.785e-4 & $\simeq3\ee$-$4\%$   & 1.085e-2& $\simeq6\ee$-$5\%$   \\
\cite{Glampedakis:2002ya} & grav - $(+)$ & 0.5 & 6 & 0.5 & 1.195e-3 & $\simeq8\ee$-$2\%$   & 1.229e-2 & $\simeq7\ee$-$2\%$  \\
\cite{Glampedakis:2002ya} & grav - $(-)$ & 0.5 & 6 & 0.1 & -1.274e-6 & $\simeq1\ee$-$1\%$  & -1.882e-5 & $\simeq2\ee$-$3\%$   \\
\cite{Glampedakis:2002ya} & grav - $(-)$ & 0.5 & 6 & 0.2 & -1.430e-6 & $\simeq5\ee$-$1\%$  & -1.973e-5&  $\simeq2\ee$-$3\%$  \\
\cite{Glampedakis:2002ya} & grav - $(-)$ & 0.5 & 6 & 0.5 & -1.126e-6 &  $\simeq8\%$   & -1.657e-5 &  $\simeq3\ee$-$2\%$\\
  \hline
  \hline
\end{tabular}
\caption{Comparison between the total (tot), horizon $(-)$ and infinity $(+)$ scalar 
and gravitational fluxes from previous works. For each quantity and 
configuration specified by the primary spin, by the eccentricity and by 
the semi-latus rectum of the secondary we show the numerical result 
obtained with our code and the relative percentage difference with the 
literature value (when available). Note that fluxes from Ref.~\cite{Warburton:2011hp} have 
a global factor 4 of difference compared to our values, due to a 
different normalization of the scalar field.}\label{tab:confsch}
\end{table}

\begin{table}[htbp!]
\begin{tabular}{ccccccc}
\hline
\hline
$\ell$ & $m$ & $n$  & $\dot{E}^{(-)}$ & $\delta\%$& $\dot{E}^{(+)}$& $\delta\%$ \\
\hline
 0& 0& 1&1.138e-8 & $\simeq1\ee$-$5\%$  &2.060e-8& $\simeq2\ee$-$4\%$\\
 0& 0& 5& 1.527e-16& $\simeq4\ee$-$7\%$ &1.926e-17& $\simeq2\ee$-$6\%$ \\
 1& 1& 1& 1.857e-8 &$\simeq3\ee$-$10\%$  & 3.256e-7& $\simeq1\ee$-$11\%$\\
 1& -1& 1&1.073e-10 &$\simeq4\ee$-$10\%$ &1.981e-9& $\simeq4\ee$-$12\%$\\
 1& 1& 5& 1.007e-14 &$\simeq2\ee$-$9\%$ &5.299e-15& $\simeq3\ee$-$8\%$\\
 1& -1& 5&4.461e-22 &$\simeq1\ee$-$6\%$ &1.993e-18& $\simeq5\ee$-$7\%$\\
 2& 2& 2&1.285e-10  &$\simeq1\ee$-$10\%$ &2.440e-8& $\simeq1\ee$-$12\%$\\
 2& -2& 2&3.955e-14 &$\simeq9\ee$-$10\%$ &2.391e-10& $\simeq4\ee$-$12\%$ \\
 2& 2& 10&1.755e-23 &$\simeq5\ee$-$5\%$ &1.842e-23& $\simeq4\ee$-$4\%$\\
 2& -2& 10&4.697e-35 &$\simeq4\ee$-$1\%$ &6.366e-30& $\simeq2\ee$-$1\%$\\
 8&  8& 10&5.882e-26 &$\simeq5\ee$-$10\%$&6.868e-18&$\simeq8\ee$-$8\%$ \\
  \hline
  \hline
\end{tabular}
\caption{Values of the scalar field energy flux at the horizon 
and at infinity that we obtained for different mode combinations,  
for a primary BH with spin $a/M=0$, and a secondary 
on eccentric orbits with $p=8M$ and $e=0.1$. For each 
quantity we show the relative errors with respect to 
the values obtained by an independent code (Courtesy of 
Niels Warburton, and derived with the code developed 
in \cite{Warburton:2011hp}).}\label{tab:conf_multipoles}
\end{table}

\end{widetext}

\section{Error estimates for the interpolation method}
\label{appendix:interpolation}

We have tested the the method used to interpolate 
energy and angular momentum fluxes, by comparing values 
of $(\dot{E}_\tn{grav,scal},\dot{L}_\tn{grav,scal})$ 
outside the numerical grid with those 
predicted by the interpolation. 
The relative differences between these two 
quantities are shown in  Table
\ref{tab:interpolationE} and \ref{tab:interpolationL}, 
for some orbital configurations.

\begin{widetext}

\begin{table}[htbp!]
\centering
\begin{tabular}{cc|cccccc}
\hline
$e$ &   $p/M$ & $\dot{E}^{int}_\text{grav}$ & $\dot{E}_\text{grav}$ & Rel. Diff. & $\dot{\bar{E}}^{int}_\text{scal}$ & $\dot{\bar{E}}_\text{scal}$ & Rel. Diff. \\
  \hline
0.1 &4  & $3.625\times10^{-3}$&  $3.631\times10^{-3}$ & 0.2\% & $1.785 \times 10^{-4}$ & $1.785\times 10^{-4}$ & 0.008\%\\
    &10 & $5.433\times10^{-5}$&  $5.132\times10^{-5}$ & 6\% & $6.865\times10^{-6}$ & $6.722\times10^{-6}$ & 2\%\\ 
\hline
0.4 &4  &  $4.838\times10^{-3}$ & $4.848\times10^{-3}$& 0.2\% & $1.964\times10^{-4}$ & $1.964\times10^{-4}$& 0.002\%\\
    &10 & $6.576 \times 10^{-5}$& $6.164\times 10^{-5}$ & 7\% &  $6.705\times10^{-6}$ & $6.528\times 10^{-6}$ & 3\%\\ 
\hline
\hline
\end{tabular}
\caption{Relative percentage difference between 
interpolated fluxes and values computed outside 
the grid of interpolation. The spin of the primary is 
fixed to $a/M=0.9$. The superscript \text{``int"} 
identifies the interpolated values.}\label{tab:interpolationE}
\end{table}

\begin{table}[htbp!]
\centering
\begin{tabular}{cc|cccccc}
\hline
$e$ &   $p/M$ & $\dot{L}^{int}_\text{grav}$ & $\dot{L}_\text{grav}$ & Rel. Diff. & $\dot{\bar{L}}^{int}_\text{scal}$ & $\dot{\bar{L}}_\text{scal}$ & Rel. Diff. \\
  \hline
0.1 &4  & $3.164\times 10^{-2}$&  $3.169\times 10^{-2}$ & 0.2\% & $6.266\times 10^{-3}$ & $6.267\times 10^{-3}$ & 0.001\%\\
    &10 & $1.659\times 10^{-3}$&  $1.635\times 10^{-3}$ & 1\% & $8.679\times 10^{-4}$ & $8.647\times 10^{-4}$ & 0.4\%\\ 
\hline
0.4 &4  &  $3.377\times 10^{-2}$& $3.383\times 10^{-2}$ & 0.2\% & $5.752\times10^{-3}$ & $5.752\times10^{-3}$& 0.0006\%\\
    &10 &  $1.576\times10^{-3}$& $1.547\times10^{-3}$ & 2\% &  $7.227\times10^{-4}$ & $7.189\times10^{-4}$ & 0.6\%\\ 
\hline
\hline
\end{tabular}
\caption{Same as Table \ref{tab:interpolationE} but for the angular momentum fluxes.}\label{tab:interpolationL}
\end{table}

\end{widetext}

\section{Waveforms}
\label{appendix:waveform}
In this appendix we provide technical details on the analytical 
templates we used to model GW signals. We follow~\cite{Barack:2003fp}, 
assuming that the waveform for the eccentric inspiral is given 
by the leading quadrupolar component, built on the seminal 
work by Peter and Matthews~\cite{Peters:1963ux}, and 
augmented by taking into account the effects of pericenter and  
Lense-Thirring precession, included with post-Newtonian corrections. 
In this work we adapted such templates to a fully-relativistic 
inspiral of equatorial eccentric geodesics described in Appendix~\ref{sec:geodesics}, which provides the semilatum 
rectum, the orbital frequencies and the eccentricity as a 
function of time through the inspiral. With these quantities 
in hand, the strain amplitude for LISA can be written as a 
sum of harmonics: 
\begin{equation}
    h_{\alpha}(t) = \sum_{n} h_{\alpha,n}(t)\quad \ , \quad  \alpha= (\tn{I},\tn{II}),
\end{equation}
where the index $\alpha$ runs on the two independent LISA 
detectors\footnote{The triangular configuration of LISA can be 
considered as a combination of two L-shaped interferometers 
with a $60$ degree angle between the arms, and rotated 
of $\pi/4$ relative each other~\cite{Cutler:1997ta}}. 
The $n$-th harmonic can be written as
\begin{equation}
    h_{\alpha,n}(t) = \frac{\sqrt{3}}{2}\left[ F_{\alpha}^{+}(t) A_n^{+}(t) + F^{\times}_{\alpha}(t) A_n^{\times}(t) \right]\ ,\label{math:strain}
\end{equation}
where $F^{+,\times}_{\alpha}$ are the detector pattern functions 
given, for the first interferometer, by
\begin{align}
F^+_\tn{I}=\frac{1+\cos^2\theta}{2}&\cos2\phi\cos2\psi-\cos\theta\sin2\phi\sin2\psi\ ,\nonumber\\
F^\times_\tn{I}=\frac{1+\cos^2\theta}{2}&\cos2\phi\sin2\psi+\cos\theta\sin2\phi\cos2\psi\ ,\nonumber
\end{align}
while for the second one $F^{+,\times}_\tn{II} =F^{+,\times}_\tn{II} (\theta, \phi - \pi/4, \psi )$. 
The plus and cross polarization then read:
\begin{align}
    h^+_{\alpha} (t) = \sum_{n} \frac{\sqrt{3}}{2}F^+_{\alpha} (t)A^+_n(t)\ , \\
      h^\times_{\alpha} (t) = \sum_{n} \frac{\sqrt{3}}{2}F^\times_{\alpha} (t)A^\times_n(t)\ .
\end{align}

The angles $(\theta, \phi, \psi)$ 
are all defined in the detector reference frame, and vary in time due 
to the LISA motion. The first two 
describe the location of the binary in the sky, while $\psi$ is 
the polarization angle. These angles can be expressed in terms of $(\theta_S,\Phi_S)$ and $(\theta_L,\Phi_L)$, which identify 
respectively the source location and the angular momentum 
$\hat{L}$ of the secondary, both in an ecliptic-based system. 
The expressions for $(\theta_S,\Phi_S)$ are given by: 
\begin{align}
&\cos\theta(t)=\frac{1}{2}\cos\theta_S-\frac{\sqrt{3}}{2}\sin\theta_S\cos[\phi_t-\phi_S]\ ,\nonumber\\
&\phi(t)=\alpha_0+\phi_t+\tan^{-1}\left[\frac{\sqrt{3}\cos\theta_S+\sin{\theta}_S\cos[\phi_t-
\phi_S]}{2\sin\theta_S\sin[\phi_t-\phi_S]}\right]\ ,
\end{align}
where $\phi_t = \bar{\phi}_0 + 2 \pi (t/T)$, $T = 1$ year 
and $(\bar{\phi}_0, \bar{\alpha}_0)$ specify the orbital 
and rotational phase of the detector when $t=0$, and are 
set to zero. The polarization angle can be expressed as 
\begin{equation}
\psi(t)=\tan^{-1}\frac{\hat{L}\cdot\hat{z}-(\hat{L}\cdot\hat{N})(\hat{z}\cdot \hat{N})}{\hat{N}\cdot(\hat{L}\times\hat{z})}\ ,
\end{equation}
with $\hat{z}\cdot\hat{N}=\cos\theta_S$ and 
\begin{align}
\hat{L}\cdot\hat{N}&=\cos\theta_L\cos\theta_S \nonumber \\
&+\sin\theta_L\sin\theta_S
\cos[\phi_L-\phi_S]\ ,\\
\hat{L}\cdot\hat{z}&=\frac{1}{2}\cos\theta_\tn{L}-\frac{\sqrt{3}}{2}\sin\theta_L\cos[\phi_t-\phi_L]\ ,\\
\hat{N}\cdot(\hat{L}\times\hat{z})&=\frac{1}{2}\sin\theta_\tn{L}\sin\theta_\tn{S}
\sin[\phi_\tn{L}-\phi_\tn{S}]\nonumber\\
&-\frac{\sqrt{3}}{2}\cos\phi_t\left[\cos\theta_\tn{L}\sin\theta_\tn{S}\sin\phi_\tn{S} \nonumber \right. \\
&\qquad\qquad\qquad\left.-\cos\theta_\tn{S}\sin\theta_\tn{L}\sin\phi_\tn{L}\right]\nonumber\\
&-\frac{\sqrt{3}}{2}\sin\phi_t\left[\cos\theta_\tn{S}\sin\theta_\tn{L}\cos\phi_\tn{L} \right. \nonumber\\
&\qquad\qquad\qquad\left.-\cos\theta_\tn{L}\sin\theta_\tn{S}\cos\phi_\tn{S}\right]\ .
\end{align}
The angles $(\theta_L, \phi_L)$ are not constant 
due to the precession of $\hat{L}$ around the MBH's spin 
direction $\hat{S}$. We can introduce then two new angles, 
$\lambda$ between $\hat{L}$ and $\hat{S}$, and $\alpha(t)$ 
measuring the precession of $\hat{L}$ around $\hat{S}$.
In this work both are set to zero, while for sake of 
simplicity we choose $(\theta_S, \phi_S,\theta_L, \phi_L) = (\pi/2, \pi/2,\pi/4, \pi/4)$. A different choice would only marginally 
change the results and the conclusions drawn in Sec.~\ref{sec:results}.
The amplitudes $A^{+,\times}_{n}$ in Eqn.~\eqref{math:strain} 
are defined as  
\begin{align}
    A_n^+ =& - [1 + (\hat{L} \cdot \hat{N})^2 ] [a_n \cos(2 \gamma) - b_n \sin(2 \gamma)] +\nonumber\\ &+[1 - (\hat{L} \cdot \hat{N})^2 ] c_n\ , \\
     A_n^\times =& 2 (\hat{L} \cdot \hat{N}) [b_n \cos(2 \gamma) + a_n \sin(2 \gamma)]\ .
\end{align}
The angle $\gamma(t)$ measures the direction of pericenter with 
respect to $\hat{x} = [-\hat{N} + \hat{L}(\hat{L} \cdot 
\hat{N})]/[1-(\hat{L} \cdot \hat{N})^2]^{1/2}$. 

In our case, in which we consider only equatorial orbits, we relate $\gamma$ to $\Psi_r$ by $\cos{(\gamma)} = \cos{(\gamma_0)} \cos{(\Psi_r)}$, where $\gamma_0$ measures the direction of the initial position of the pericenter with 
respect to $\hat{x}$, and $\Psi_r$ is the angle in the orbital plane defined in \eqref{eq:single_phases}.
We chose $\gamma_0 = \pi/4$. The coefficients $(a_n,b_n,c_n)$ are then given by 
\begin{align}
a_n = & -n \mathcal{A}\big[J_{n-2}(n e) - 2 e J_{n-1}(n e)+(2/n)J_n(n e) \nonumber\\
 & + 2e J_{n+1}(ne)-J_{n+2}(ne)\big]\cos[n\Phi(t)]\ ,\\
 b_n =& -n \mathcal{A} (1-e^2)^{1/2}[J_{n-2}(ne)-2J_{n}(n e) \nonumber\\
 & +J_{n+2}(ne)]\sin[n\Phi(t)]\ ,\\
 c_n =& 2 \mathcal{A} J_n(ne) \cos[n\Phi(t)]\ , 
\end{align}
where $J_{n}$ is the Bessel function of the first kind, 
$\mathcal{A}= (2 \pi \nu M )^{2/3} \mu/ D $, with 
$2 \pi \nu = \dd\Phi/\dd t $ and $D$ being the source 
luminosity distance. 
In this work we fix $\Phi = \Psi_\phi$, such that 
$2 \pi \nu = \Omega_\phi$.

\bibliographystyle{utphys}
\bibliography{Ref}

\end{document}